\pdfoutput=1
\documentclass[11pt,twoside,a4paper,cmspaper,final,collab]{cms-tdr}
\def\svnVersion{3cecdbf}\def\svnDate{2024/03/26}\def\cmsCernNoTag{CERN-EP-2023-216}\def\cmsCernDate{\today}\def\cmsMessage{Published in the Journal of High Energy Physics as \href{http://dx.doi.org/10.1007/JHEP03(2024)134}{\doi{10.1007/JHEP03(2024)134}.}}
\begin{document}\cmsNoteHeader{EXO-21-012}

\newcommand{\Semi}{\ensuremath{\Pell\PGn\PQq\PQq}\xspace} 
\newcommand{\Full}{\ensuremath{2\Pell2\PGn}\xspace} 
\newcommand{\Hs}{\ensuremath{\mathrm{s}}\xspace} 
\newcommand{\Jet}{\ensuremath{\mathrm{j}}}
\newcommand{\lmax}{\ensuremath{\Pell_{\text{max}}}\xspace}
\newcommand{\lmin}{\ensuremath{\Pell_{\text{min}}}\xspace}
\newcommand{\ptww}{\ensuremath{\pt^{\PW\PW}}\xspace}
\newcommand{\hww}{\ensuremath{\PH\to\PWp\PWm}\xspace}
\newcommand{\mTlminmet}{\ensuremath{\mT^{\lmin, \ptmiss}}\xspace}
\newcommand{\mTllmet}{\ensuremath{\mT^{\Pell\Pell, \ptmiss}}\xspace}
\newcommand{\mTlmet}{\ensuremath{\mT^{\Pell, \ptmiss}}\xspace}
\newcommand{\ptveclmin}{\ensuremath{\ptvec^{\lmin}}\xspace}
\newcommand{\pTmissprojPF}{\ensuremath{\pt^{\text{miss}, \text{PF proj}}}\xspace}   
\newcommand{\pTmissprojTk}{\ensuremath{\pt^{\text{miss},\text{track proj}}}\xspace}  
\newcommand{\pTmissproj}{\ensuremath{\pt^{\text{miss}, \text{proj}}}\xspace}
\newcommand{\mll}{\ensuremath{m_{\Pell\Pell}}\xspace}
\newcommand{\mjj}{\ensuremath{m_{\Jet\Jet}}\xspace}
\newcommand{\mljjmet}{\ensuremath{m_{\Pell\Jet\Jet\ptmiss}}\xspace}
\newcommand{\ptlmax}{\ensuremath{\pt^{\lmax}}\xspace}
\newcommand{\ptlmin}{\ensuremath{\pt^{\lmin}}\xspace}
\newcommand{\ptll}{\ensuremath{\pt^{\Pell \Pell}}}
\newcommand{\ptjj}{\ensuremath{\pt^{\Jet \Jet}}}
\newcommand{\ptljj}{\ensuremath{\pt^{\Pell \Jet \Jet}}}
\newcommand{\delRll}{\ensuremath{\Delta R_{\Pell\Pell}}\xspace}
\newcommand{\delRljj}{\ensuremath{\Delta R_{\Pell,\Jet \Jet}}\xspace}
\newcommand{\delPljj}{\ensuremath{\Delta \phi_{\Pell,\Jet \Jet}}\xspace}
\newcommand{\delPljjmet}{\ensuremath{\Delta \phi_{\Pell\Jet\Jet, \ptvecmiss}}\xspace}
\newcommand{\delPlmet}{\ensuremath{\Delta \phi_{\Pell,\ptvecmiss}}\xspace}
\newcommand{\ttst}{\ensuremath{\PQt}\xspace\text{production}\xspace}
\newcommand{\wjets}{\ensuremath{\PW{+}}jets\xspace}
\newcommand{\ptl}{\ensuremath{\pt^{\Pell}}}
\newcommand{\ptjo}{\ensuremath{\pt^{\Jet_{1}}}}
\newcommand{\ptjt}{\ensuremath{\pt^{\Jet_{2}}}}
\newcommand{\MADDM} {\textsc{MadDM}\xspace}

\newlength\cmsTabSkip\setlength{\cmsTabSkip}{1ex}
\providecommand{\cmsTable}[1]{\resizebox{\textwidth}{!}{#1}}

\cmsNoteHeader{EXO-21-012}
\title{Search for dark matter particles in \texorpdfstring{\PWp\PWm}{W+W-} events with transverse momentum imbalance in proton-proton collisions at \texorpdfstring{$\sqrt{s} = 13\TeV$}{sqrt(s) = 13 TeV}}

\author{The CMS Collaboration}
\date{\today}

\abstract{
A search for dark matter particles is performed using events with a pair of \PW bosons and large missing transverse momentum. Candidate events are selected by requiring one or two leptons ($\Pell =$ electrons or muons). The analysis is based on proton-proton collision data collected at a center-of-mass energy of 13\TeV by the CMS experiment at the LHC and corresponding to an integrated luminosity of 138\fbinv. No significant excess over the expected standard model background is observed in the \Semi and \Full final states of the $\PWp\PWm$ boson pair. Limits are set on dark matter production in the context of a simplified dark Higgs model, with a dark Higgs boson mass above the $\PWp\PWm$ mass threshold. The dark matter phase space is probed in the mass range 100--300\GeV, extending the scope of previous searches. Current exclusion limits are improved in the range of dark Higgs masses from 160 to 250\GeV, for a dark matter mass of 200\GeV.
}

\hypersetup{
pdfauthor={CMS Collaboration},
pdftitle={Search for dark matter particles produced in W+W- events with transverse momentum imbalance in proton-proton collisions at sqrt(s) = 13 TeV},
pdfsubject={CMS},
pdfkeywords={CMS, dark matter search, dark Higgs model}}

\maketitle 

\section{Introduction}\label{sec:intro}

Astrophysical and cosmological observations~\cite{Gaitskell:2004gd, Trimble:1987ee, Porter:2011nv, ref:DMevi} provide abundant evidence that dark matter (DM) exists. 
However, this evidence is based only on gravitational interactions, and whether DM has nongravitational interactions with standard model (SM)
particles is still one of the major questions of fundamental physics. Many theories of new physics at the scale of electroweak (EW) symmetry
breaking~\cite{ref:DMNPBSM} propose viable candidates for DM and are able 
to accommodate the observed relic density of DM particles in the universe~\cite{ref:relic,ref:Planck}.
Compelling contenders for DM are weakly interacting massive particles (WIMPs, denoted as $\chi$)~\cite{ref:WIMPs}, 
which could be produced at high-energy colliders such as the Large Hadron Collider (LHC) at CERN. 
A favored DM signature in collider searches consists of one or more SM particles, \PX, that are produced and detected,
recoiling against a pair of noninteracting DM particles that escape detection, resulting in missing transverse momentum (\ptmiss).
Previous searches at the LHC took \PX to be a light quark producing a jet~\cite{ref:monojet,ref:monojet1}, 
a heavy-flavor (bottom or top) quark~\cite{ref:monob2,ref:monob}, a photon~\cite{ref:monog,ref:monog1}, 
or a \PW, \PZ, or Higgs boson~\cite{ref:monoW,Sirunyan:2017qfc,ref:monojet1,ref:monoW1,Sirunyan_2020}. 

\begin{figure}[!h]
  \centering
  \includegraphics[width=0.38\textwidth]{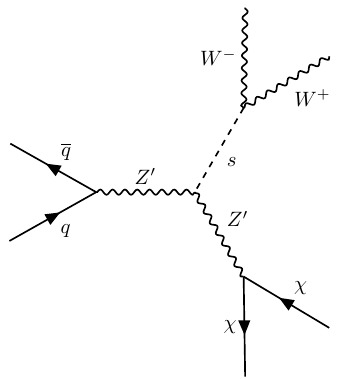}
  \hspace{1cm}
  \includegraphics[width=0.475\textwidth]{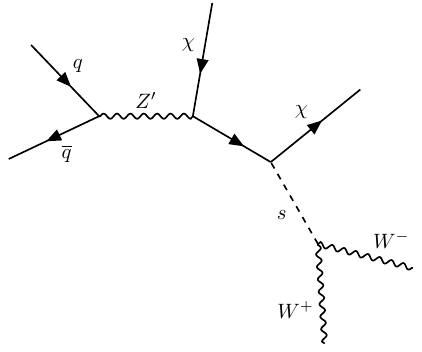}
  \caption{The dominant Born-level Feynman diagrams for the benchmark signal model considered in this paper:
    $\qqbar \to \PZpr \to \Hs \chi \chi$, and $\Hs \to \PWp\PWm$.}
  \label{fig:models}
\end{figure}

An approach to probe DM at the LHC is based on a scenario in which the DM particle acquires mass through its interaction with a dark Higgs field~\cite{Duerr_2017}.
The signal model used as a benchmark in this search posits a Majorana DM particle that transforms under a new U(1) local gauge symmetry 
yielding an additional massive spin-1 vector boson \PZpr and a new physical dark Higgs boson \Hs which is a singlet of the SM gauge groups.
The \PZpr mediator could be responsible for establishing thermal equilibrium between the visible and the dark sector in the early universe 
and could provide the annihilation and creation processes that set the DM relic abundance via thermal freeze-out. 
Current results constrain the parameter space in which the DM particles can acquire their relic abundance 
from direct annihilation into SM final states~\cite{WIMPSsafer}. 
This limitation can be significantly relaxed if the DM particle is not isolated as the lightest state in the dark sector.
The dark Higgs boson can be lighter than the DM particle $\chi$, 
in which case the observed relic abundance can readily be achieved through $\chi\chi \to \Hs\Hs$ annihilation. 
Even if \Hs is not strictly lighter than the DM candidate, the limitation can be evaded by the $\chi\chi \to \Hs\PZpr$ 
annihilation channel if $m_{\Hs}$ is low enough~\cite{Bell_2016}.
The dominant Born-level Feynman diagrams contributing to the $\Hs{+}\chi\chi$ signature 
are shown in Fig.~\ref{fig:models} and involve a \PZpr boson or $\chi$ intermediate state.
The relevant model parameters are the DM mass $m_{\chi}$, the \PZpr mass 
$m_{\PZpr}$, the dark Higgs boson mass $m_{\Hs}$, the \PZpr couplings to quarks ($g_{\Pq}$) and to DM particles ($g_{\chi}$), 
and the mixing angle ($\theta$) between the SM and the dark Higgs bosons.

In this paper, a DM search is described using data recorded with the CMS detector at $\sqrt{s} = 13\TeV$ in 2016--2018, corresponding to an integrated luminosity of 138\fbinv.
The \PX plus \ptmiss signature is targeted, where \PX
refers to the dark Higgs boson decaying to a $\PWp\PWm$ vector boson pair. 
For this paper, the signatures in which the \PW bosons both decay to charged leptons ($\Pell =$ electrons or muons) and neutrinos (\Full),
or in which one decays to a charged lepton and a neutrino and the other decays to a pair of quarks (\Semi), are considered.
For $m_{\Hs} >  160\GeV$, the $\PWp\PWm$ channel has the largest branching fraction~\cite{ATL-PHYS-PUB-2019-032}.
This is the first search addressing such a dark Higgs model performed by the CMS Collaboration.
Limits on this model have been published by the ATLAS Collaboration at $\sqrt{s} = 13\TeV$ using $\PW\PW$ and $\PZ\PZ$ channels, 
with the vector bosons decaying hadronically~\cite{ref:ATLAS-DarkHiggs_1} or decaying to one lepton and jets~\cite{ref:ATLAS-DarkHiggs_2}.

This paper is organized as follows. A brief introduction to the CMS detector is given in Section~\ref{sec:detector}.
The data and simulated event samples are described in Section~\ref{sec:datasim}. 
The event reconstruction is detailed in Section~\ref{sec:reco}. The analysis strategy and the event selections are detailed
in Sections~\ref{sec:anal} and~\ref{sec:eventselection}, respectively. Section~\ref{sec:bkgestim} describes the background 
estimation and Section~\ref{sec:systematics} the systematic uncertainties. Finally, the results are presented in 
Section~\ref{sec:results}, and the final summary is given in Section~\ref{sec:summary}.
Tabulated results are provided in the HEPData record for this analysis~\cite{hepdata}.

\section{The CMS detector}\label{sec:detector}

The central feature of the CMS apparatus is a superconducting solenoid of 
6\unit{m} internal diameter, providing a magnetic field of 3.8\unit{T}. 
Within the solenoid volume are a silicon pixel and strip tracker, a lead 
tungsten crystal electromagnetic calorimeter (ECAL), and a brass and 
scintillator hadron calorimeter (HCAL), each composed of a barrel and 
two endcap sections. Forward calorimeters extend the pseudorapidity ($\eta$) coverage 
provided by the barrel and endcap detectors. Muons are detected in 
gas-ionization chambers embedded in the steel flux-return yoke outside the solenoid. 
The first level of the CMS trigger system~\cite{CMSTrigger}, composed of custom hardware processors, 
is designed to select the most interesting events within a time interval of less than 4\unit{$\mu\mathrm{s}$}, 
using information from the calorimeters and muon detectors,  with an 
output rate of up to 100\unit{kHz}. The high-level trigger processor farm 
further reduces the event rate to about 1\unit{kHz} before data storage.
A more detailed description of the CMS detector, together with a definition of the
coordinate system used and the relevant kinematic variables, can be found 
in Ref.~\cite{Chatrchyan:2008zzk}.

\section{Data and simulated samples}\label{sec:datasim}

The proton-proton ($\Pp\Pp$) collision data used in this search were collected at $\sqrt{s} = 13\TeV$
during 2016, 2017, and 2018, with integrated luminosities of 36.3, 41.5, and 59.8\fbinv, 
respectively~\cite{CMS-LUM-17-003,CMS-PAS-LUM-17-004,CMS-PAS-LUM-18-002}. The average number of multiple $\Pp\Pp$ 
interactions per bunch crossing (pileup) is approximately 23 for the 2016 data, and 32 for the 2017--2018 data.

Events are stored if they satisfy the selection criteria of online triggers requiring one or two 
leptons with a minimum transverse momentum (\pt) requirement. The lowest \pt thresholds for the double-lepton 
triggers are 23\GeV for the leading lepton (\lmax) and 12\GeV for the next-to-leading lepton (\lmin). The 
single-lepton triggers in the 2016 data set have \pt thresholds of 25\GeV for $\abs{\eta} < 2.1$ and
27\GeV for $2.1 < \abs{\eta} < 2.5$ for electrons, and of 24\GeV for muons.
In the 2017 data set, the thresholds are increased to 35 for electrons and 27\GeV for muons, while in the 2018 data set
they are reduced to 32 and 24\GeV, respectively. The trigger efficiency is measured using \PZ{+}jets events and is larger than 90\%
for both electrons and muons over the given $\eta$ range.

Monte Carlo (MC) simulated events are used for modeling both signal and background processes. Event samples for 2016, 2017, 
and 2018 data sets are simulated separately to account for the differences in the detector and pileup conditions. Different parton 
distribution functions (PDFs) and underlying event (UE) tunes are used for the different simulated data sets. In all simulations, 
the \PYTHIA~\cite{Sjostrand:2014zea} 8.226 (8.230) library is used for parton showering, hadronization, and the 
UE simulation, using the CUETP8M1~\cite{Khachatryan:2015pea} (CP5~\cite{Sirunyan:2019dfx}) tune for 2016 (2017--2018). 
Similarly, the NNPDF 3.0~\cite{Ball:2013hta,Ball:2011uy} PDF set is used for the 2016 data set, 
while the NNPDF 3.1~\cite{Ball:2017nwa} set is used for the 2017--2018 data sets. 

Signal samples are generated at leading order (LO) in perturbative quantum chromodynamics (QCD) using the \MGvATNLO~v2.6.5 
generator~\cite{Alwall_2014}, applying the \PYTHIA CP5 tune for all three years. 
A scan in the ($m_{\Hs}$, $m_{\PZpr}$, $m_{\chi}$) parameter space is performed on 672 points between $160 < m_{\Hs} < 400\GeV$,
$200 < m_{\PZpr} < 2500\GeV$, and $100 < m_{\chi} < 300\GeV$, while the other model parameters are fixed to 
$g_{\chi} = 1.00$, $g_{\Pq} = 0.25$, and $\sin{\theta} = 0.01$.
The $\PZpr$ and \Hs boson widths vary slightly depending on the selected parameters and are confirmed to be below 1\% of their respective masses. 
The chosen values for these parameters correspond to those recommended as a benchmark in the LHC Dark Matter Working Group (LHC DM WG)~\cite{Abercrombie_2020, Duerr_2017}. 

Continuum $\PWp\PWm$ production background via \qqbar annihilation is generated with \POWHEG~v2
at next-to-LO (NLO) precision~\cite{Melia:2011tj}, while $\Pg\Pg \to \PW\PW$ events are generated at LO
using \MCFM v7.0~\cite{Campbell:2011bn,Campbell:2015qma}.
The simulated $\qqbar \to \PW\PW$ events are reweighted to match the \pt distribution of the $\PW\PW$-system (\ptww) from the \pt-resummed calculation at
next-to-NLO plus next-to-next-to-leading logarithmic precision~\cite{Meade:2014fc,Jaiswal:2014yba},
and to account for the higher-order EW effects~\cite{gieseke2014vectorboson}. The LO $\Pg\Pg \to \PW\PW$
cross section, obtained from \MCFM, is scaled to next-to-NLO precision via a $K$ factor of 1.4~\cite{Caola:2015rqy}. 
The different Higgs boson production modes are generated with
\POWHEG~v2~\cite{Nason:2004rx,Frixione:2007vw,Alioli:2010xd,Bagnaschi:2011tu,Nason:2009ai,Luisoni:2013kna,Hartanto:2015uka}
and the \hww decay with \textsc{JHUgen}~\cite{Bolognesi:2012mm}.
Single top quark, \ttbar, $\PW\PZ$, and $\PW\PGg^*$ background processes are generated at NLO with \POWHEG v2.
To further improve the modelling of \ttbar production, the simulated samples are reweighted based on the \pt of each top quark~\cite{Czakon_2017}.
Production of a single \PW boson in association with jets is simulated at LO using \MGvATNLO v2.2.2 and v2.6.5 for 2016 and 2017--2018 samples respectively.
The description of the \wjets process is improved by scaling the events to kinematic distributions computed at NLO EW precision~\cite{Lindert:2017}.
The QCD $K$ factors are derived from NLO samples as a function of the generated \pt of the \PW boson.
Variations of up to 12\% of the simulated event yields due to different generator versions 
and settings across the sample years are compensated for with an additional $K$ factor depending
on the reconstructed dijet mass, and treated as uncertainties.
Drell--Yan (DY) production of $\PZ/\PGg^{*}$, and other multiboson processes, 
such as $\PW\PGg$, $\PZ\PZ$, and $\PV\PV\PV$ ($\PV = \PW$ or \PZ), are generated at NLO using \MGvATNLO.
All samples are normalized to the latest available theoretical cross sections with at least NLO precision~\cite{Campbell:2011bn,Melnikov:2006kv}. 
For all the processes, the detector response is simulated using the \GEANTfour toolkit~\cite{Agostinelli:2002hh}.

\section{Event reconstruction}\label{sec:reco}

The primary vertex (PV) is taken to be the vertex corresponding to the hardest scattering in the event, 
evaluated using tracking information alone, as described in Section 9.4.1 of Ref.~\cite{CMS-TDR-15-02}.

A particle-flow (PF) algorithm~\cite{CMS-PRF-14-001} aims to reconstruct and identify individual particles in an event 
with an optimized combination of information from the various elements of the CMS detector.
The energy of photons is obtained from the ECAL measurement.
The energy of charged hadrons is determined from a combination of their momentum measured in the tracker
and the matching ECAL and HCAL energy deposits, corrected for the response function of the calorimeters to hadronic showers.
The energy of neutral hadrons is obtained from the corresponding corrected ECAL and HCAL energies.

Electrons are reconstructed using information from the pixel detector, the silicon strip tracker, and the ECAL 
in the interval $\abs{\eta} < 2.5$. The track is required to be consistent with 
originating from the PV. The electron momentum is estimated by combining the energy 
measurement in the ECAL, the momentum measurement in the tracker, and the energy sum of all bremsstrahlung photons 
spatially compatible with the electron track. The efficiency to reconstruct and identify electrons ranges between 60--80\% 
depending on the electron \pt and $\eta$. The momentum resolution for electrons with $\pt\approx 45\GeV$, 
measured in $\PZ \to \Pe\Pe$ decays, ranges from 1.7--4.5\%, depending on the $\eta$ region and on the bremsstrahlung energy 
emitted by the electron as it traverses the material in front of the ECAL~\cite{Khachatryan:2015hwa}. 

Muons are reconstructed in the interval $\abs{\eta} < 2.4$ using the information from the muon chambers 
and the silicon tracker, and they are required to be consistent with the reconstructed PV. The efficiency 
to reconstruct and identify muons is greater than 96\%. The relative \pt resolution for muons 
with \pt up to 100\GeV is 1\% in the barrel and 3\% in the endcaps~\cite{Sirunyan:2018fpa,Sirunyan:2019yvv}.

Both electrons and muons are required to be isolated. 
In the case of an electron, the scalar \pt sum of other PF candidates within a cone of size 
$\Delta R = \sqrt{\smash[b]{(\Delta\eta)^2+(\Delta\phi)^2}} < 0.3$ (where $\phi$ is the azimuthal angle in radians) 
around its direction is required to be less than 6\% of the electron \pt. 
Only those PF candidates associated with the PV are included in the isolation sum.
In case of a muon, the \pt sum is evaluated for PF candidates in a cone $\Delta R < 0.4$ 
and is required to be less than 15\% of the muon \pt.         
       
Jets in each event are clustered from the neutral and charged PF candidates associated with the 
PV, using the anti-\kt algorithm~\cite{Cacciari:2008gp, Cacciari:2011ma} with a distance parameter of $R = 0.4$.
The expected average contribution from pileup is subtracted from the reconstructed jet energy~\cite{CMS-PRF-14-001}.
Additional selection criteria are applied to each jet with $\pt > 15\GeV$ and $\abs{\eta}<4.7$ 
to remove those potentially dominated by anomalous contributions from 
various subdetector components or reconstruction failures~\cite{CMS-PAS-JME-16-003}. 
For the 2017 data sets, jets in the range $2.5<\abs{\eta}<3.0$ are excluded to eliminate spurious jets caused by the detector noise.
The jet energy resolution typically amounts to 15\% at 10\GeV, 8\% at 100\GeV, and 4\% at 1\TeV~\cite{Khachatryan_2017}.

The identification of jets containing hadrons with bottom (\Pb) quarks is referred to as \Pb tagging. For
each reconstructed jet, a \Pb tagging score is calculated through a multivariate analysis of jet properties
using a fully-connected deep neural network based algorithm, DeepCSV~\cite{Sirunyan:2017ezt}. Jets
are considered \Pb tagged if the DeepCSV \Pb tag output score is above a threshold set to achieve ${\approx}80\%$
efficiency for \Pb quark jets in \ttbar events. 
For this threshold, the probability of misidentifying a jet as \Pb quark jet in \ttbar events is ${\approx}11\%$ for light-flavor or gluon jets and ${\approx}41\%$ for charm quark jets.

The vector (\ptvecmiss), with  magnitude \ptmiss, is defined as the negative vector \pt sum 
of all the PF candidates in an event~\cite{Sirunyan:2019kia}, weighted by their estimated
probability to originate from the primary interaction vertex. The pileup-per-particle identification 
algorithm~\cite{Bertolini:2014bba} is employed to calculate this probability.
Energy scale corrections to the reconstructed jets are taken into account in this computation.

\section{Analysis strategy}\label{sec:anal}

This analysis exploits the dependence of the kinematic properties of the final-state objects 
(the decay products of the $\PWp\PWm$ boson pair and a substantial amount of \ptmiss from DM) on the three parameters
$m_{\Hs}$, $m_{\PZpr}$ and $m_{\chi}$, which enables the separation of signal events 
from the main background events. 

For the \Full channel, 
the observable chosen to test the dark Higgs model is the transverse mass of the \lmin plus \ptmiss system, 
the distribution of which is more sensitive to the predicted dark Higgs signal than other observables based on the kinematic properties of the lepton or \ptmiss.
This transverse mass is defined as:
\begin{linenomath}
\begin{equation}
  \mTlminmet = \sqrt{2 \ptlmin \ptmiss \left[1 - \cos\Delta\phi(\ptveclmin, \ptvecmiss)\right]},
  \label{eq:mTlminmet}
\end{equation}
\end{linenomath}
where \ptlmin is the magnitude of the \pt vector of the \lmin candidate, 
and $\Delta\phi(\ptveclmin, \ptvecmiss)$ is the azimuthal angle between \ptveclmin, and the \ptvecmiss.
The signal is extracted from a two-dimensional profiled maximum likelihood fit to the invariant mass of the dilepton system, 
\mll and the \mTlminmet observable with a signal-plus-background hypothesis using the \textsc{RooStats} Project tools~\cite{Moneta:2010pm}. 
The asymptotic approximation of the profile likelihood ratio~\cite{Cowan2010AsymptoticFF} is used as a
test statistic.

For the \Semi channel,
because of the relatively low expected cross section of the signal processes and the large irreducible \wjets 
background, the variables that show most promising separation of signal and background are
combined in a boosted decision tree (BDT) to produce a single discriminator. 
From the initial set of 23 candidate variables, a subset of 13 variables is selected by comparing intermediate BDTs 
in terms of background rejection and signal selection efficiency. 
An overview of all selected variables and their descriptions is given in Table~\ref{tab:BDT_var}.
The resulting discriminator returns values between $-1$, the background-like extreme, and 1, the most signal-like.
It has been cross-checked that events with values at the lower end are predominantly background, while signal
events accumulate at higher values. The final analysis results are obtained from a fit to the
binned BDT discriminator distribution, as described in Section~\ref{sec:results}.

\begin{table}[htbp]
  \centering
    \topcaption{Summary of all selected variables considered in the BDT for the \Semi channel. \label{tab:BDT_var}}
      \begin{tabular} {lp{10cm}}
        \hline
        Variable                                                 & Definition\\ 
        \hline
        \ptjj                                                    & \pt of the vectorial sum of the \PW candidate jets\\
        \ptljj                                                   & \pt of the vectorial sum of the visible particles\\
        \ptmiss                                                  & Magnitude of the missing transverse momentum vector\\
        $\Delta \eta_{\Pell,\Jet\Jet}$ and \delPljj              & $\Delta \eta$ and $\Delta \phi$ between the lepton and the dijet system\\
        $\Delta \eta_{\Jet,\Jet}$ and $\Delta \phi_{\Jet,\Jet}$  & $\Delta \eta$ and $\Delta \phi$ between the \PW candidate jets\\

        $\abs{ \eta_{\Pell} }$                                   & The absolute value of the lepton pseudorapidity\\
        \delPlmet                                                & $\Delta \phi$ between the lepton and \ptvecmiss\\

        \delPljjmet                                              & $\Delta \phi$ between the vectorial sum of the visible particles and \ptvecmiss\\
        $\min(\ptl,\ptjt)/\ptmiss$                       & Minimum of the lepton \pt and the next-to-leading \PW candidate jet \pt, divided by \ptmiss\\ 
        $\mathrm{max}(\ptl,\ptjo)/\ptmiss$                       & Maximum of the lepton \pt and the leading \PW candidate jet \pt, divided by \ptmiss\\
        $\mathrm{max}(\ptl,\ptjo)/\mljjmet$                      & Maximum of the lepton \pt and the leading \PW candidate jet \pt, divided by the invariant mass of the system of all visible particles and \ptvecmiss, which is taken to be massless\\ 
        \hline
      \end{tabular}
\end{table}

\section{Event selection}\label{sec:eventselection}

\subsection{The \texorpdfstring{\Full}{2l2nu} channel} 

The main feature of the $\Hs \to \PWp\PWm$ decay in the \Full channel is the presence of two oppositely
charged and isolated leptons with relatively large \pt, recoiling against \ptvecmiss.
The analysis imposes a set of requirements on kinematic and topological quantities aimed at defining
a phase space enriched in $\Hs \to \PWp\PWm$ events.

In addition to the requirement of opposite charge, the leptons are required to be well-identified, isolated, and have different flavors.
In order to reduce the contamination from the DY process, final states with same-flavor leptons are excluded in this analysis.
The \ptlmax is required to be greater than 25\GeV and the \ptlmin has to be greater than 20\GeV, 
to ensure a good reconstruction and identification efficiency. 
The dilepton system is required to have a minimum invariant mass \mll of 20\GeV in order to eliminate low-mass DY.
The transverse momentum of the lepton pair \ptll must be greater than 30\GeV, 
to reduce background contributions from leptons that do not originate from the PV (nonprompt leptons). 
To minimize the impact of multilepton contributions from $\PW\PZ$ and $\PZ\PZ$ backgrounds, 
events that include a third lepton, which is subjected to less strict selection criteria 
compared to the lepton identification used for the two high-\pt leptons, are rejected if its \pt is larger than 10\GeV.
Events with one or more \PQb-tagged jets with transverse momentum $\pt^{\PQb}>20\GeV$ are rejected. This selection reduces
the background from top quark production by ${\approx}86\%$ while losing less than 10\% of signal events.

To further suppress DY background, a minimum value of the transverse mass of the dilepton system plus \ptmiss
is required, $\mTllmet > 50\GeV$, and a selection on the quantity \pTmissproj~\cite{Sirunyan:2018egh}
is introduced. This quantity is defined as the component of \ptvecmiss in the direction of the nearest lepton 
if the lepton is situated within the azimuthal angular cone of $\pm{\pi/2}$ from the \ptvecmiss direction,
or the \ptmiss otherwise.
A selection using this variable efficiently rejects $\PZ/\PGg^* \to \Pell\Pell$ background events in which the \ptvecmiss 
is preferentially aligned with leptons. Since the \ptmiss resolution is degraded by pileup, a quantity is defined 
as the smaller of the two \pTmissproj values: one based on all the PF candidates in the event (\pTmissprojPF),
and the other one based only on 
the reconstructed tracks originating from the PV (\pTmissprojTk).
This quantity is required to be larger than 20\GeV.
These selection requirements remove ${\approx}95\%$ of the DY background and less than 5\% of the signal.

\begin{figure}
\centering
\includegraphics[width=0.55\textwidth]{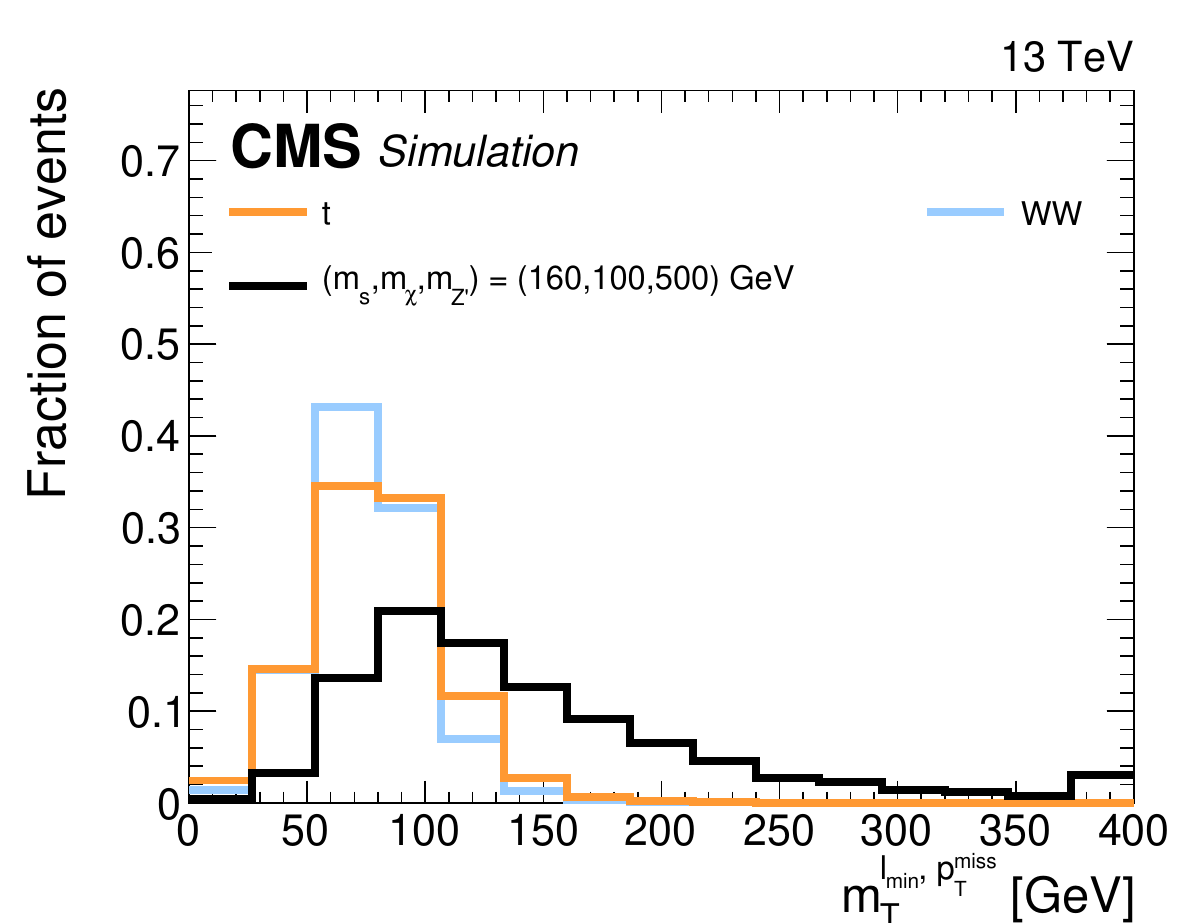}
\caption{Normalized \mTlminmet distribution in the \Full channel for a signal with $m_{\Hs} = 160\GeV$, $m_{\chi} = 100\GeV$, $m_{\PZpr} = 500\GeV$ (black), after the event selection criteria are applied.
Predictions for the two main backgrounds of the analysis, $\PW\PW$ and \ttst, are shown as blue and yellow solid lines respectively. The last bin includes the overflow. The \ttst sample consists of approximately 79\% \PQt quark pair production and 21\% single \PQt processes.}
\label{fig:mt_shapes}
\end{figure}

A summary of the event selection is shown in Table~\ref{tab:event_sel_em}, 
and the \mTlminmet distribution for signal and leading backgrounds is shown in Fig.~\ref{fig:mt_shapes}. 

\begin{table}[htbp]
  \centering
    \topcaption{Summary of the event selection criteria in the \Full channel. \label{tab:event_sel_em}}
      \begin{tabular} {lcc}
        \hline
        Quantity                                     & Selection \\ 
        \hline
        Number of leptons                            & 2         \\
        Lepton flavors                               & $\Pe\Pgm$ \\
        Lepton charges                               & Opposite  \\
        Additional leptons                           & 0         \\
        \ptlmax                                      & $>$25\GeV\\
        \ptlmin                                      & $>$20\GeV\\
        \mll                                         & $>$20\GeV\\ 
        \ptll                                        & $>$30\GeV\\ 
        \ptmiss                                      & $>$20\GeV\\
        $\min(\pTmissprojPF,\pTmissprojTk)$          & $>$20\GeV\\
        \mTllmet                                     & $>$50\GeV\\
        \delRll                                      & $<$2.5   \\
        Number of \PQb-tagged jets                   & 0         \\
        \hline
      \end{tabular}
\end{table}

To account for the dependence of the kinematic properties of the final-state objects on the mass of the dark Higgs boson,
the events that pass the event selection described above are categorized into three signal regions (SRs).
The split is based on a proxy of the Lorentz boost of the dark Higgs boson recoiling against the DM system by 
considering the distance between the two leptons in the ($\eta$,$\phi$) plane, \delRll.
An optimized division is achieved in SR1: $\delRll < 1.0$ (high-boost); SR2: $ 1.0 < \delRll < 1.5$ (medium-boost); 
and SR3: $1.5 < \delRll < 2.5$ (low-boost) by carefully balancing the signal efficiency and background rejection.
Approximately 33\% of the background is rejected by selecting events with $\delRll < 2.5$, with a negligible effect on the signal.

\subsection{The \texorpdfstring{\Semi}{lnuqq} channel}

The main feature in the \Semi channel of the $\Hs \to \PWp\PWm$ decay is the presence of one well-reconstructed and isolated lepton
and at least two jets recoiling against \ptvecmiss. Different lepton preselections are used for each data set to comply with the varying
trigger selection thresholds; they are listed in Table~\ref{tab:leppt}. 
Events containing a second loosely-defined lepton with $\pt > 10\GeV$ are vetoed.

\begin{table}[htbp]
  \centering
  \topcaption{
    Selection criteria for the leptons for 2016--2018 data in the \Semi channel. The \pt thresholds are chosen to be equal to the corresponding single-lepton trigger threshold.}
    \begin{tabular}{ccc}
      \hline
      Year & Electron \pt  & Muon \pt \\
      \hline
      2016 & $>$25\GeV & $>$24\GeV \\
      2017 & $>$35\GeV & $>$27\GeV \\
      2018 & $>$32\GeV & $>$24\GeV \\
      \hline
    \end{tabular}
  \label{tab:leppt}
\end{table}

When more than two jets are available, the pair having an invariant mass
closest to the \PW boson mass are selected as \PW candidate jets.
Both \PW candidate jets are required to be within the tracker acceptance region ($\abs{\eta} < 2.4$) 
to avoid the high multiplicity of background particles produced in the high-$\abs{\eta}$ regions.
Events with jets that do not originate from a \PW boson are suppressed by requiring $65 < \mjj < 105\GeV$.
The contribution of top quark background events to the SR is further reduced by removing
events with at least one b-tagged jet, with $\pt > 20\GeV$, that is not tagged as a \PW candidate jet.

All background events can be further suppressed by comparing the kinematic variables of the visible particles and \ptmiss.
The presence of DM candidates in an event results in a large \ptmiss well separated from visible particles as compared to SM processes. 
Thus, we impose the requirements $\ptmiss > 60\GeV$ and $\mTlmet > 80\GeV$ where
\begin{linenomath}
\begin{equation}\label{eq:mtlmet}
    \mTlmet = \sqrt{2\,\pt^{\Pell} \ptmiss [1 - \cos (\delPlmet)]}.
\end{equation}
\end{linenomath}
These two kinematic requirements are especially effective in decreasing the amount of nonprompt-lepton and DY background.
Another feature of the signal events is that the visible particles tend to cluster azimuthally in the direction opposite to \ptvecmiss.
For these reasons, \delRljj is required to be smaller than 3.0, \delPljj is 
required to be below 1.8, \delPljjmet must be above 2.0, and \ptljj must be above 60\GeV.

These additional topological requirements remove ${\approx}97\%$ of the background 
(${\approx}97\%$ for \wjets, ${\approx}94\%$ for \ttst, and ${\approx}99\%$ for nonprompt-lepton events) while keeping between 60--80\% of signal events. 
An optimal differentiation of signal and background events is accomplished using
a BDT for which these variables are inputs.
A summary of the event selection criteria for the \Semi channel is shown in Table~\ref{tab:event_sel_semi}.

\begin{table}[htbp]
  \centering
    \topcaption{Summary of the event selection criteria for the \Semi channel. \label{tab:event_sel_semi}}
      \begin{tabular} {lcc}
        \hline
        Quantity                           & Selection \\
        \hline
        Number of leptons                  & 1         \\
        Additional leptons                 & 0         \\
        Number of jets                     & $\geq$2  \\
        Non-\PW-candidate \PQb-tagged jets & 0         \\
        \mjj                               & $>$65\GeV, $<$105\GeV \\
        \ptmiss                            & $>$60\GeV\\
        \ptljj                             & $>$60\GeV\\
        \mTlmet                            & $>$80\GeV\\
        \delRljj                           & $<$3.0   \\
        \delPljj                           & $<$1.8   \\
        \delPljjmet                        & $>$2.0   \\[1ex]
        \hline
      \end{tabular}
\end{table}

\section{Background estimation}\label{sec:bkgestim}

All background processes, except those with nonprompt leptons, 
are modeled using MC simulations.
The nonprompt-lepton background is estimated from data by examining events selected with less stringent lepton selection criteria.
These loose leptons have a relative high probability to originate from jets that have been incorrectly categorised.
By measuring the rate of these misidentified loose leptons erroneously passing the SR criteria 
and the efficiency of correctly reconstructing and identifying a lepton, 
the nonprompt contribution to the SR can be estimated.
Details of this method are given in Ref.~\cite{Sirunyan:2018egh}. 
The validity of this background estimation is checked in a validation region. 
For the \Full selection, this region is defined by requiring two same-sign leptons 
while keeping the other selection criteria the same as those listed in Table~\ref{tab:event_sel_em}.
In the \Semi case, the requirements $\ptmiss < 30\GeV$ and $\mTlmet < 30\GeV$
replace the corresponding selections of Table~\ref{tab:event_sel_semi}, 
exploiting the fact that events with a nonprompt lepton do not necessarily have a neutrino, 
resulting in a relatively small \ptmiss. 

The normalizations of the most important background processes are determined from the observed data in certain control regions (CRs). 
The main backgrounds consist of the $\PWp\PWm$ and DY processes for the \Full channel, 
\wjets events for the \Semi channel, and \ttst for both.
One CR is defined for each process: 
\begin{itemize}
\item The \ttst CR is defined by requiring the presence of at least one \PQb-tagged jet, thereby inverting the \PQb-veto. 
\item In the \Full channel, the $\PWp\PWm$ CR is defined by inverting the \delRll selection, \ie, $\delRll > 2.5$. 
\item In the \Full channel, the DY CR is defined by inverting the \mTllmet requirement, \ie, $\mTllmet < 50\GeV$.
\item In the \Semi channel, the \wjets CR is defined by inverting the \mjj selection, \ie, $\mjj < 65$ or $> 105\GeV$. 
\end{itemize}
The event yields in these four CRs are fitted simultaneously with the SRs, 
letting the $\PWp\PWm$, DY, \wjets and \ttst normalizations float freely in all CRs and SRs.
A possible signal contribution in the CRs is also taken into account.

\section{Systematic uncertainties}\label{sec:systematics}

Experimental and theoretical sources of systematic uncertainty are described in this section. Effects  
from  experimental uncertainties are studied by scaling and/or smearing the reconstructed objects in simulation and 
propagating the effect to the kinematic variables used in the analysis. 

There are several sources of experimental systematic uncertainties, including the lepton reconstruction and identification efficiencies, 
the lepton momentum scales, the jet energy scale and resolution, the b tagging efficiency for 
\PQb~quark jets, the mistag rate for light-flavor quark and gluon jets, the modeling of \ptmiss and of 
pileup in the simulation, the background contributions, and the integrated luminosity. All experimental sources 
are treated as uncertainties that either change the shapes of the distributions 
or scale the normalizations of the individual signal and background processes.
In the simultaneous fit, 
the different sources of uncertainty are each represented by single independent nuisance parameters.
When combining the uncertainties across processes and data sets, 
they are treated as correlated or uncorrelated as appropriate.

Lepton efficiency uncertainties are evaluated as functions of the lepton \pt and $\eta$, using the tag-and-probe method.
The impact of the trigger efficiency uncertainty is less than 1 (2)\% overall in the \Full (\Semi) analysis,
while the uncertainties in the reconstruction and identification efficiency 
cause shape and normalization changes of $\approx$1.0\% ($\approx$1.0\%) for electrons and $\approx$2.0 ($\approx$0.5)\% for muons. 
Changes in lepton momentum scale are at the level of 0.2 (0.4)\% for muons and $\approx$0.6--1.0 ($\approx$0.1--0.8)\% for electrons. 
Those uncertainties are propagated to \ptmiss resulting in a total variation of 1--10\%. 
For the changes in the jet energy scale, the impact on the normalization is $\approx$1\% for the \Full search,
while for the \Semi case, the impact can be as large as 10\%.

Experimental uncertainties in the estimation of the nonprompt-lepton background are also taken into account. The nonprompt background is 
affected by the shape uncertainties arising from the dependence on the flavor composition of the jets misidentified as leptons. 
These shape uncertainties amount to $\approx$5--10\%~\cite{Sirunyan:2018egh}. 
Statistical fluctuations in the data sets from which the scale factors are determined can be sizable in the \Semi analysis, 
with a 2--40\% variation for electrons and 4--25\% for muons. For the \Full channel, the trigger conditions are less tight, 
resulting in larger data sets for estimating efficiencies with negligible statistical uncertainties.
In addition, a 30\% normalization uncertainty is determined from a closure test performed with simulated samples.

The modeling of pileup depends on the total inelastic $\Pp\Pp$ cross section~\cite{Sirunyan:2018nqx}. 
The pileup uncertainty is evaluated by varying this cross section up and down by 5\%.
The uncertainty in the integrated luminosity measurement is 1.2, 2.3, and 2.5\% for the
2016, 2017, and 2018 data sets, respectively~\cite{CMS-LUM-17-003,CMS-PAS-LUM-17-004,CMS-PAS-LUM-18-002},
combining to an overall 1.6\% uncertainty across the three years;
it contributes directly to the normalization of the MC samples.
Luminosity uncertainties are not considered for the top quark, \wjets, $\PWp\PWm$, and the DY background processes
as their normalizations are determined by the fit to the data. 

Several theoretical uncertainties are pertinent for all simulated event samples. Uncertainties in this
category arise from the choice of PDFs and missing higher-order corrections in the perturbative
expansion of the simulated cross sections. The PDF uncertainties are estimated, following the PDF4LHC recommendations~\cite{Butterworth:2015oua}, 
comparing the value obtained using the set of MC replicas provided by the NNPDF collaboration and PDFs with varied \alpS values. 
The estimated uncertainties from missing higher-order corrections in the perturbative QCD expansion are given by the bin-by-bin 
difference between the nominal and alternative templates, which are constructed from simulated 
events, where renormalization and factorization scales are shifted up and down by a factor of two. 
The parton showering scale uncertainties are estimated by varying the initial-state and 
final-state radiation scales separately by a factor two. An overall 1.5\% UE uncertainty is applied to all the simulated samples 
in order to account for fluctuations obtained from alternative MC simulations with UE tune variations~\cite{CMS:2022uhn}. 

Theoretical systematic uncertainties specific to individual background
processes are also considered. A 15\% uncertainty is applied to the relative fraction of the
gluon-induced component of the $\PWp\PWm$ background~\cite{Caola:2016trd}, while for the $\qqbar \to \PW\PW$ component, the applied 
\ptww corrections are varied by changing the renormalization and factorization scales. 
For the \ttbar component in the \ttst background samples, the entire \pt correction weight 
(as mentioned in Section~\ref{sec:datasim}) is treated as the uncertainty.
Furthermore, the entire EW NLO correction to the \wjets samples is considered as an uncertainty, 
while variations in the QCD NLO corrections are estimated from statistical fluctuations in the sample sizes, 
and by varying the jet $\abs{\eta}$ selections in the $K$ factor estimations, resulting in 1 and 2\% uncertainties, respectively.
Differences in year-by-year setup for the \wjets simulation are addressed
with an additional simulation-based nuisance parameter and amount to a $\approx$12\% variation.

Finally, uncertainties arising from the limited statistical precision of the simulation
are included for each bin of the discriminant distributions in each independent category~\cite{BARLOW1993219}.
In the \Semi channel, these typically add an uncertainty of $\approx$20\% in the highest BDT bins.
For the \Full channel, the values vary greatly among the different analysis regions, typical values are $\approx$1--40\% in SR1, $\approx$1--20\% in SR2 and $\approx$1--10\% in SR3.

\section{Results}\label{sec:results}

Distributions from the \Full and \Semi channels are fitted jointly, and the signal
is obtained from the results of the fit. These distributions correspond to both SRs and CRs.
 
\begin{figure}[!hbtp]
\centering
\includegraphics[width=0.49\textwidth]{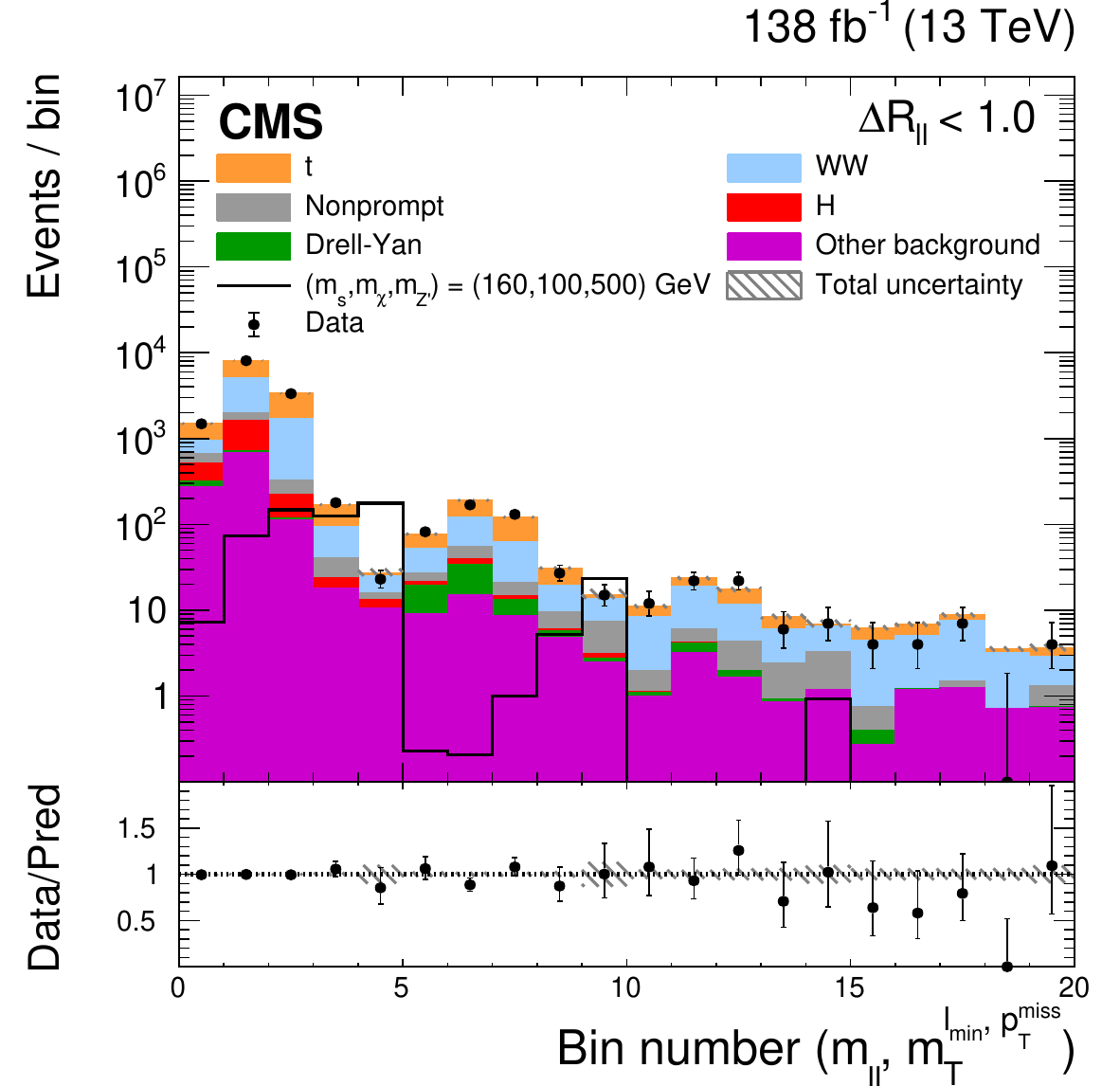} 
\includegraphics[width=0.49\textwidth]{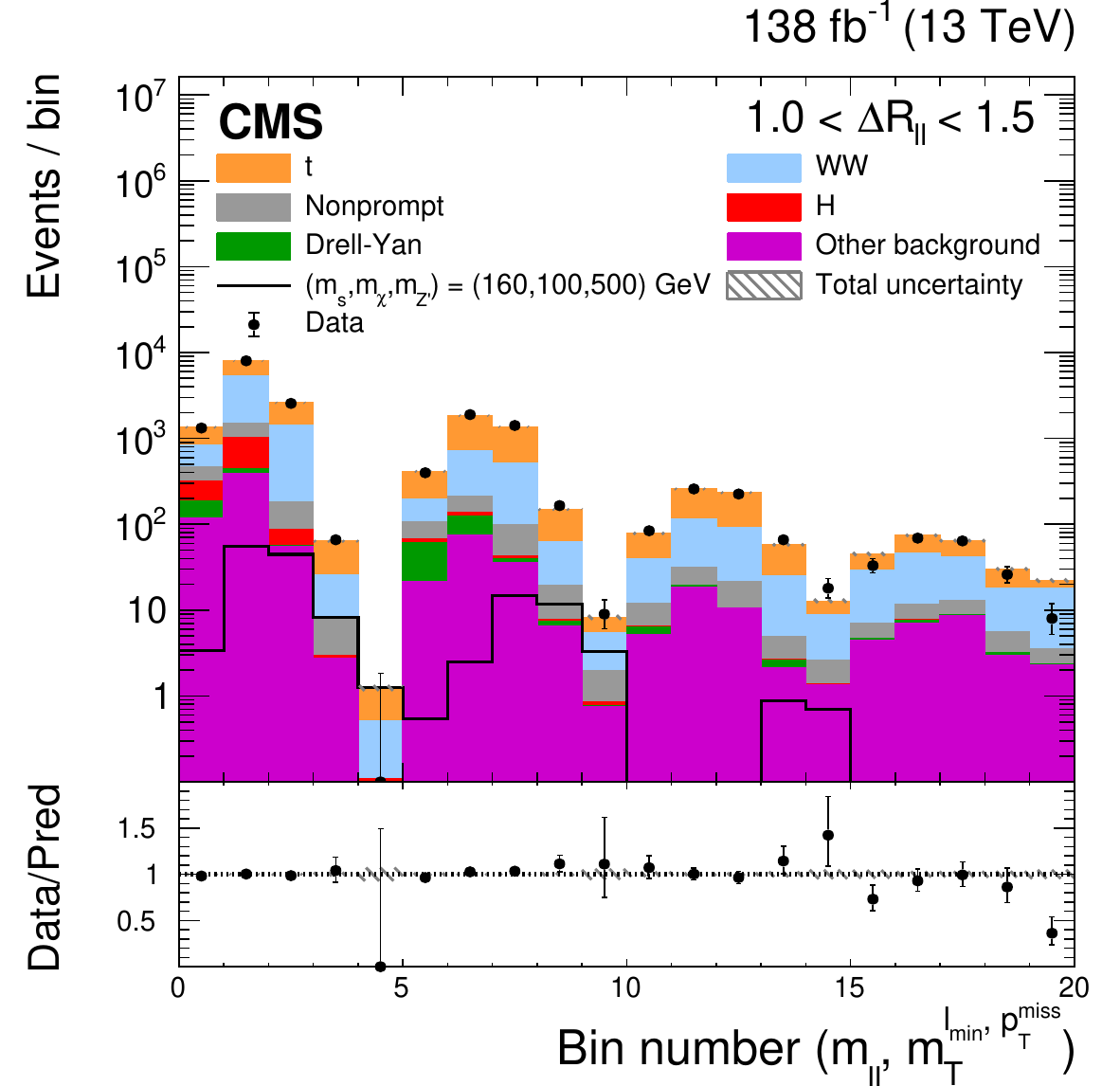} \\
\includegraphics[width=0.49\textwidth]{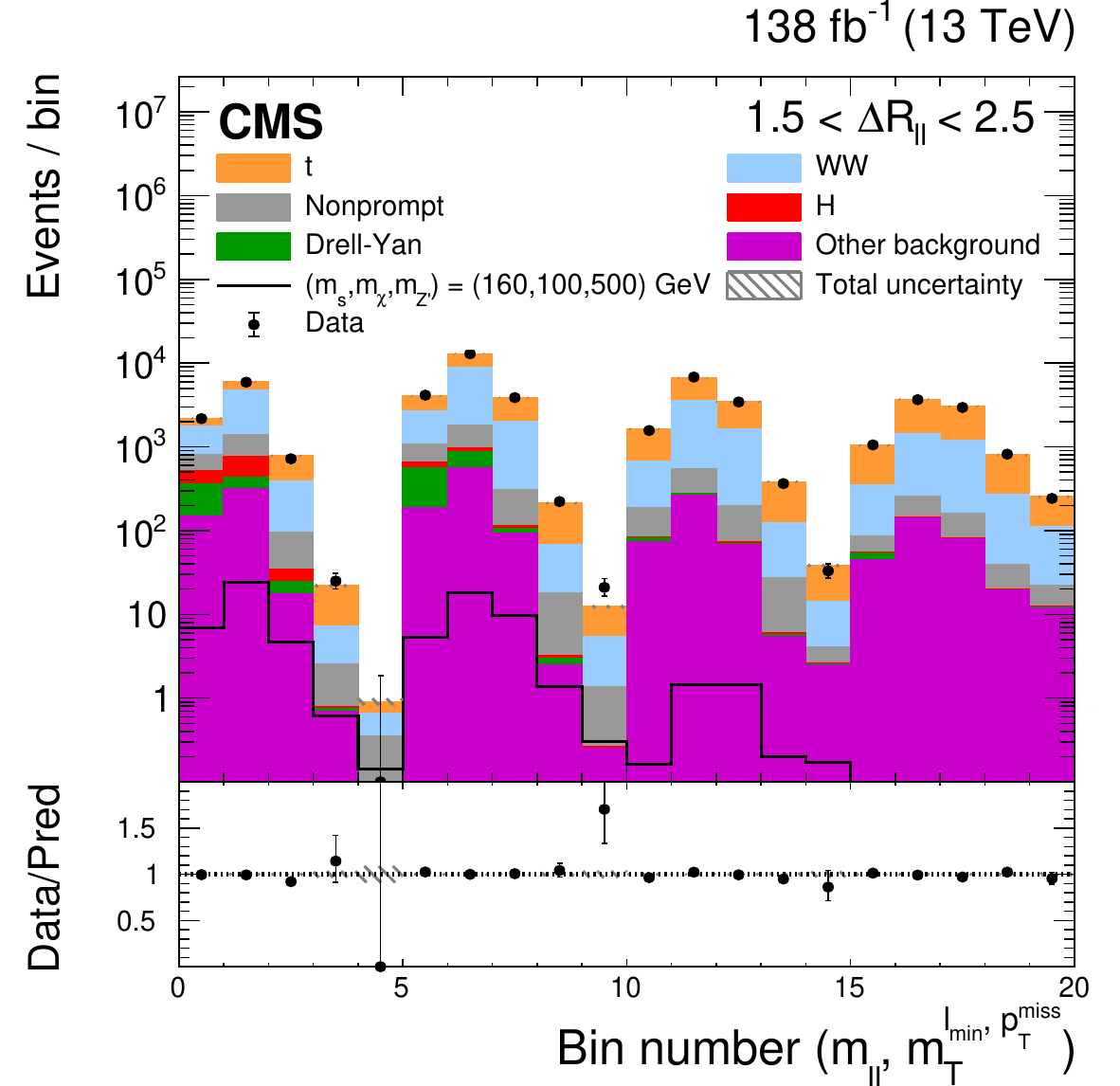} 
\caption{\label{fig:2D_DH_df} Unrolled (\mll,\mTlminmet) post-fit distributions in the \Full channel 
  in a given \delRll region SR1 (upper left), SR2 (upper right), and SR3 (lower), for the full data set.
  The histogram bins are spaced uniformly. Each group of five bins (from left to right) corresponds to the \mTlminmet distribution
  in a \mll region, placed in ascending order.
  The black line indicates the signal prediction for $m_{\Hs} = 160\GeV$, $m_{\chi} = 100\GeV$, $m_{\PZpr} = 500\GeV$.
  In the lower panel of each plot, the ratio between the data and the background prediction is shown.}
\end{figure}

For the \Full channel, the event yields in the CRs are recorded in single-bin histograms.
Kinematic information from the SR is contained in two-dimensional distributions of \mll and \mTlminmet.
The \mll binning is [12, 60, 90, 120, $\infty$)\GeV is chosen by taking into account the significance
as a function of \mll for each dark Higgs boson mass.
The \mTlminmet binning is chosen by optimizing the sensitivity to the different signal mass point distributions, 
subject to the requirements that bins that are more sensitive contain at least 10 events, 
and bins that are less sensitive contain at least 1 event in each data period: 
[0, 50, 90, 130, 160, $\infty$), [0, 50, 90, 130, 170, $\infty$), and [0, 50, 90, 130, 180, $\infty$)\GeV for 2016, 2017, and 2018 data sets respectively.
This strategy provides flexibility to the analysis, and allows the hypothetical signals to freely populate
the (\delRll,\mll,\mTlminmet) phase space according to the kinematic properties of the different dark Higgs boson mass points
while establishing a uniform procedure for modeling  background.
In Fig.~\ref{fig:2D_DH_df}, the (\mll,\mTlminmet) post-fit distributions for the full integrated luminosity of 138\fbinv are unrolled into contiguous 1-dimensional slices showing the \mTlminmet distributions for separate regions of \mll. 
Post-fit expected background and data yields for the three SRs are listed in Table~\ref{tab:SR_em_postfit}.

\begin{table}[h!]
\centering
\topcaption{
  Data and background yields for each analysis region in the \Full channel.
  Central values and uncertainties for the background contributions are the post-fit values.
  For the signal prediction from simulation, with the associated uncertainties, values are given for a sample with
  $m_{\Hs} = 160\GeV$, $m_{\chi} = 100\GeV$, $m_{\PZpr} = 500\GeV$.}
\cmsTable{
\begin{tabular}{lr@{}lr@{}lr@{}lr@{}lr@{}lr@{}l}
\hline
Process                       &\multicolumn{2}{c}{$\PW\PW$ CR} &  \multicolumn{2}{c}{DY CR} & \multicolumn{2}{c}{\PQt quark CR} &        \multicolumn{2}{c}{SR1} &    \multicolumn{2}{c}{SR2} &        \multicolumn{2}{c}{SR3} \\
\hline
\PQt quark                    & $   18000 $ & $\pm      510$ & $     953 $ & $\pm       38$ & $  347800 $ & $\pm     1300$ & $    5480 $ & $\pm      170$ & $    7090 $ & $\pm      210$ & $   21250 $ & $\pm      600$ \rule{0pt}{2ex}\\
Nonprompt                     & $    3160 $ & $\pm      370$ & $     247 $ & $\pm       30$ & $    9600 $ & $\pm     1200$ & $     738 $ & $\pm       92$ & $    1050 $ & $\pm      120$ & $    3500 $ & $\pm      410$ \\
DY                            & $     240 $ & $\pm       11$ & $    2171 $ & $\pm       74$ & $     517 $ & $\pm       35$ & $   112.6 $ & $\pm      7.0$ & $     211 $ & $\pm       11$ & $    1042 $ & $\pm       48$ \\
$\PW\PW$                      & $   19250 $ & $\pm      660$ & $     517 $ & $\pm       36$ & $    3330 $ & $\pm      230$ & $    4980 $ & $\pm      200$ & $    6760 $ & $\pm      240$ & $   22480 $ & $\pm      710$ \\
$\PV\PZ$                      & $    28.3 $ & $\pm      1.0$ & $    27.0 $ & $\pm      1.4$ & $    29.4 $ & $\pm      1.8$ & $   24.29 $ & $\pm     0.97$ & $    28.8 $ & $\pm      1.0$ & $    63.8 $ & $\pm      2.4$ \\
$\PV\gamma$ + $\PV\gamma^{*}$ & $    1580 $ & $\pm      170$ & $     152 $ & $\pm       13$ & $     644 $ & $\pm       82$ & $    1088 $ & $\pm       96$ & $     703 $ & $\pm       74$ & $    1870 $ & $\pm      200$ \\
$\PV\PV\PV$                   & $    74.3 $ & $\pm      2.8$ & $    8.67 $ & $\pm     0.38$ & $    63.2 $ & $\pm      3.8$ & $   25.73 $ & $\pm     1.00$ & $    28.9 $ & $\pm      1.1$ & $    77.5 $ & $\pm      2.9$ \\
\PH                           & $    56.4 $ & $\pm      1.5$ & $   109.1 $ & $\pm      3.9$ & $     516 $ & $\pm       25$ & $    1202 $ & $\pm       29$ & $     759 $ & $\pm       19$ & $     712 $ & $\pm       18$ \\
Signal                        & $   3.409 $ & $\pm    0.097$ & $   3.119 $ & $\pm    0.081$ & $    70.0 $ & $\pm      2.4$ & $   561.8 $ & $\pm      6.5$ & $   147.1 $ & $\pm      1.7$ & $   74.30 $ & $\pm     0.76$ \\[\cmsTabSkip]
Total bgnd                    & $   42390 $ & $\pm      190$ & $    4184 $ & $\pm       64$ & $  362520 $ & $\pm      600$ & $   13649 $ & $\pm       87$ & $   16634 $ & $\pm       84$ & $   50990 $ & $\pm      180$ \\[\cmsTabSkip]
Data                          & $   42397 $ & $            $ & $    4183 $ & $            $ & $  362508 $ & $            $ & $   13627 $ & $            $ & $   16681 $ & $            $ & $   50918 $ & $            $ \\
\hline
\end{tabular}
} 
\label{tab:SR_em_postfit}
\end{table}

\begin{figure}[!hbtp]
\centering
\includegraphics[width=0.49\textwidth]{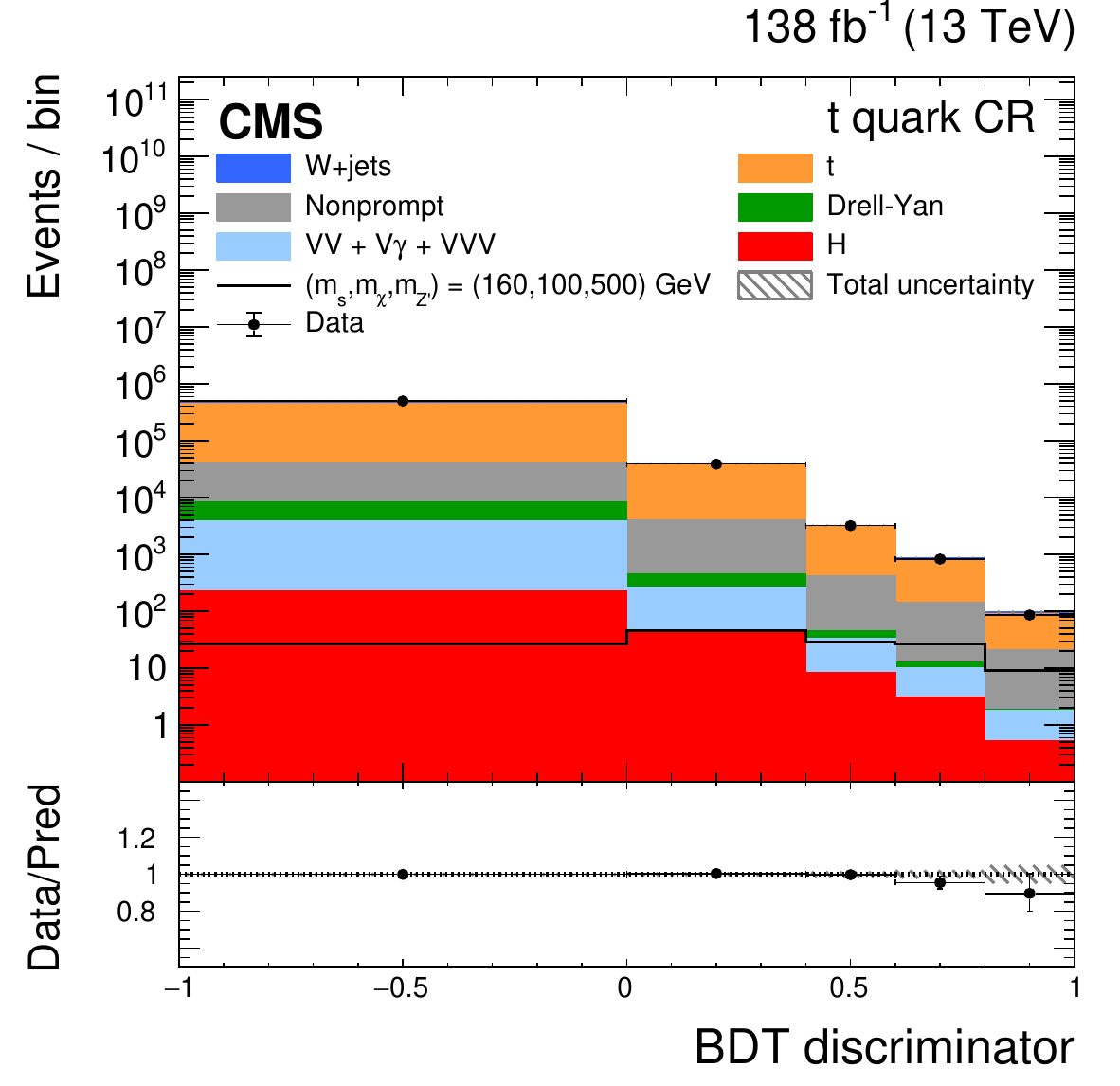} 
\includegraphics[width=0.49\textwidth]{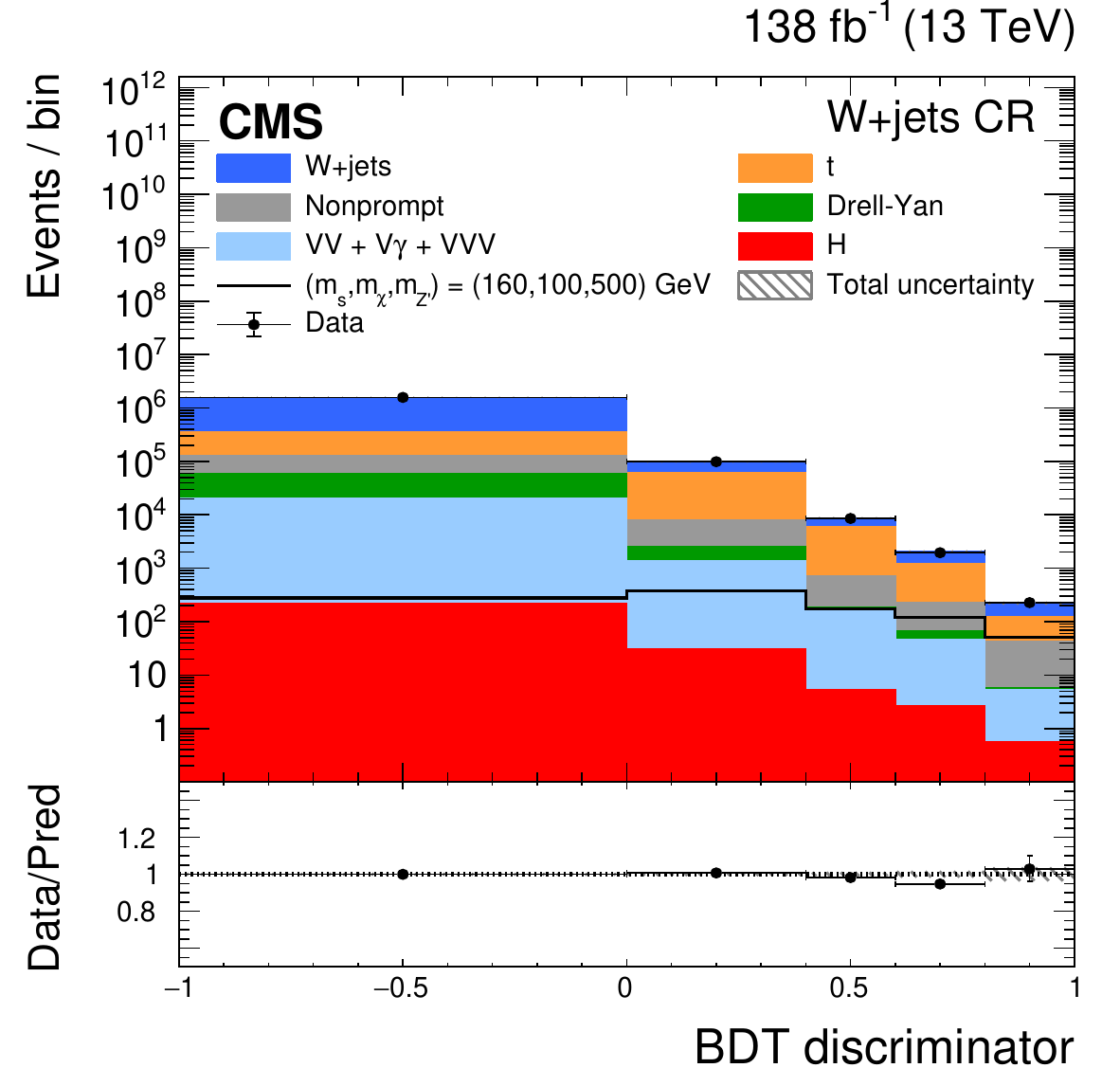} \\
\includegraphics[width=0.49\textwidth]{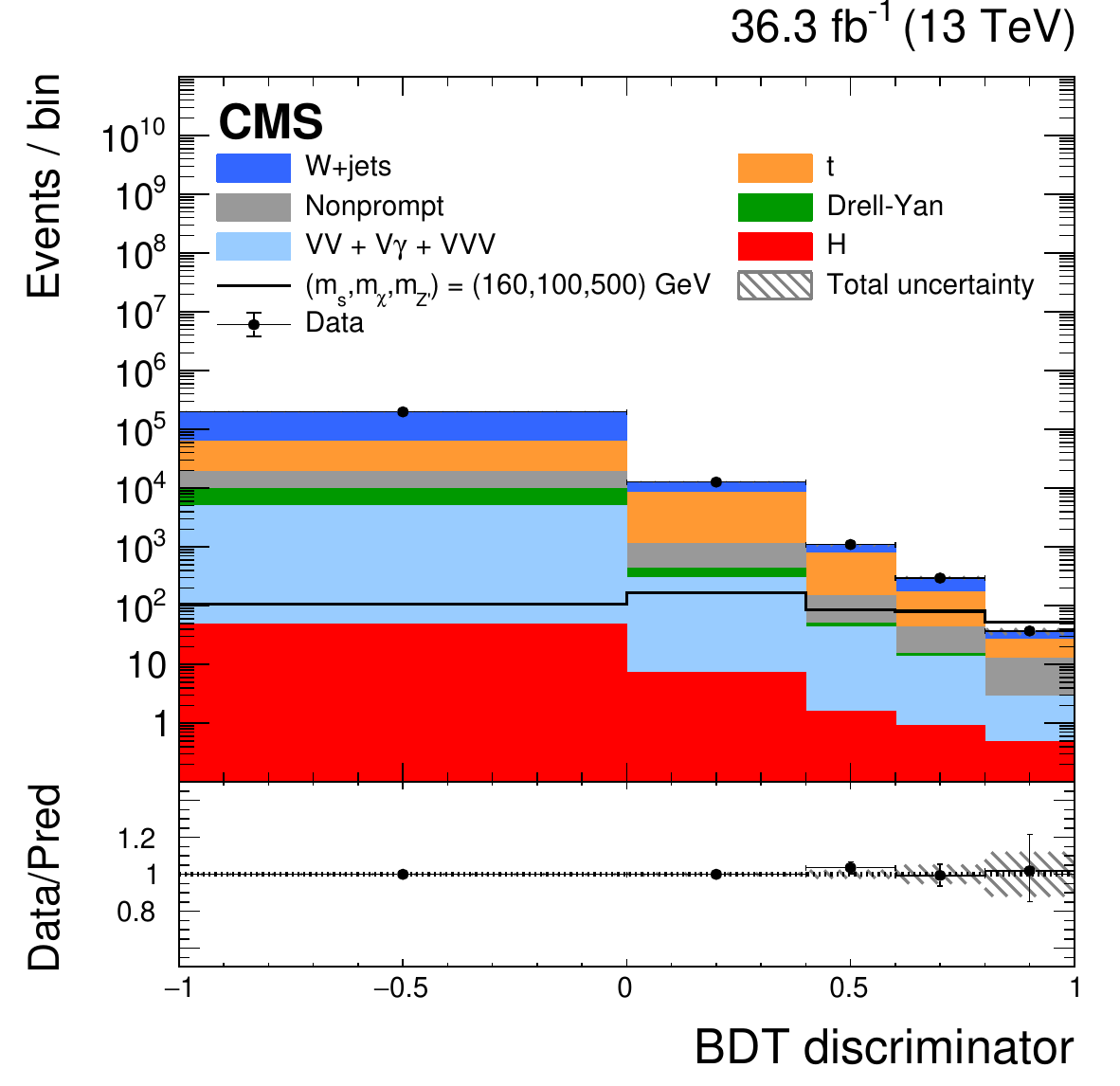} 
\includegraphics[width=0.49\textwidth]{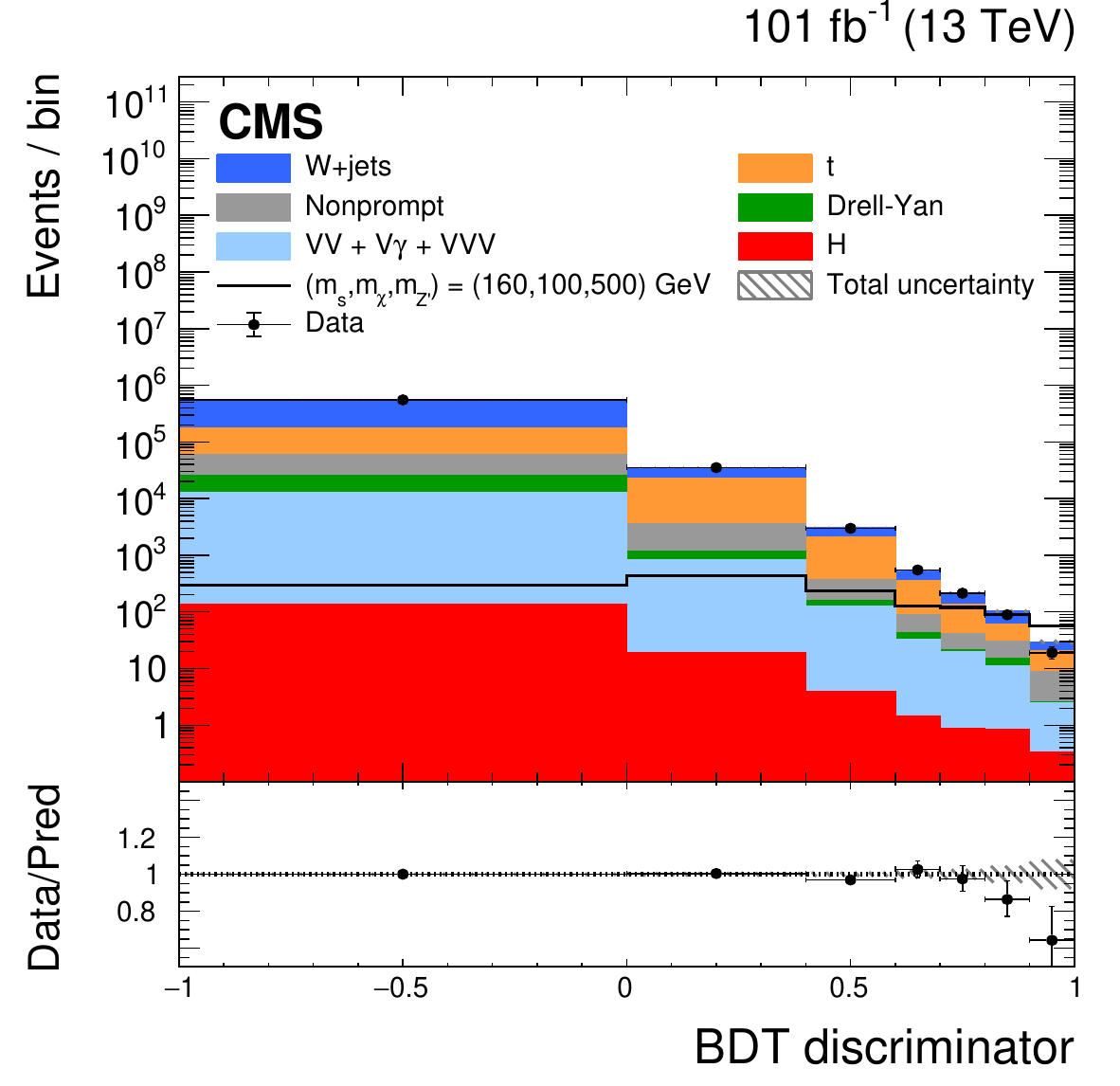} 
\caption{\label{fig:BDT_DH_df} Post-fit BDT distributions in the \Semi channel for the full data set
  in the \PQt quark CR (upper left) and \wjets CR (upper right).
  The SR of the 2016 data set (lower left) and the 2017--2018 data set (lower right).
  The black line indicates the signal prediction with $m_{\Hs} = 160\GeV$, $m_{\chi} = 100\GeV$, $m_{\PZpr} = 500\GeV$.
  In the lower panel of each plot, the ratio between the data and the background prediction is shown.}
\end{figure}

In the \Semi channel, the signal and background information is contained in BDT distributions
for the SRs and CRs.  
The binning is set by considering the bin-to-bin significance of dark Higgs boson mass points and by
requiring the expected yield to be greater than 10 events in every bin.
A coarser binning is obtained for the CRs and also for the 2016 SR ([$-1$, 0, 0.4, 0.6, 0.8, 1])
while a finer binning is obtained for the 2017--2018 SRs ([$-1$, 0, 0.4, 0.6, 0.7, 0.8, 0.9, 1]).
The post-fit distributions of the SRs and CRs are shown in Fig.~\ref{fig:BDT_DH_df} for
the two CRs (upper two plots corresponding to 138\fbinv), the 2016 SR
(lower left plot, 36.3\fbinv) and the combined 2017--2018 SRs (lower right plot, 101\fbinv).
In the \PQt quark CR and the 2017--2018 SR distributions shown in Fig.~\ref{fig:BDT_DH_df},
a downward slope is visible in the last two bins of the distribution of 
the ratio of the observed to predicted events.
This apparent trend is found to be an artifact of the chosen binning.
Table~\ref{tab:semi_pf_yields} shows the post-fit expected yields of the background processes and
the pre-fit signal for the sample with $m_{\Hs} = 160\GeV$, $m_{\chi} = 100\GeV$, $m_{\PZpr} = 500\GeV$,
when the BDT discriminator values are above 0.6.

\begin{table}[h!]
\centering
\topcaption{Data and background yields for the \Semi channel with a BDT discriminator score above 0.6. 
  Central values and uncertainties for the background contributions are the post-fit values.
  For the signal prediction from simulation, with the associated uncertainties, values are given for a sample with
  $m_{\Hs} = 160\GeV$, $m_{\chi} = 100\GeV$, $m_{\PZpr} = 500\GeV$.}
\begin{tabular}{lr@{}lr@{}lr@{}l}
\hline
Process                       &  \multicolumn{2}{c}{\wjets CR} & \multicolumn{2}{c}{\PQt quark CR} &       \multicolumn{2}{c}{SR} \\
\hline
\wjets                        & $     916 $ & $\pm       42$ & $    52.0 $ & $\pm      5.0$ & $     461 $ & $\pm       27$ \rule{0pt}{2ex}\\
\PQt quark                    & $    1035 $ & $\pm       28$ & $     742 $ & $\pm       24$ & $     531 $ & $\pm       15$ \\
Nonprompt                     & $     201 $ & $\pm       27$ & $     142 $ & $\pm       23$ & $     124 $ & $\pm       18$ \\
DY                            & $    19.9 $ & $\pm      4.9$ & $    2.89 $ & $\pm     0.51$ & $    17.2 $ & $\pm      4.0$ \\
$\PW\PW$                      & $    23.3 $ & $\pm      2.7$ & $    4.70 $ & $\pm     0.51$ & $    44.6 $ & $\pm      3.4$ \\
$\PV\PZ$                      & $   0.190 $ & $\pm    0.015$ & $   0.682 $ & $\pm    0.055$ & $   0.371 $ & $\pm    0.024$ \\
$\PV\gamma$ + $\PV\gamma^{*}$ & $      22 $ & $\pm       12$ & $    0.78 $ & $\pm     0.13$ & $    13.4 $ & $\pm      4.4$ \\
$\PV\PV\PV$                   & $    3.81 $ & $\pm     0.16$ & $    1.62 $ & $\pm     0.13$ & $    8.91 $ & $\pm     0.60$ \\
\PH                           & $    3.09 $ & $\pm     0.15$ & $    3.61 $ & $\pm     0.18$ & $    4.71 $ & $\pm     0.21$ \\
Signal                        & $   172.2 $ & $\pm      3.3$ & $    35.5 $ & $\pm      2.1$ & $   528.1 $ & $\pm      9.3$ \\[\cmsTabSkip]
Total background              & $    2225 $ & $\pm       39$ & $     950 $ & $\pm       22$ & $    1205 $ & $\pm       29$ \\[\cmsTabSkip]
Data                          & $    2179 $ & $            $ & $     917 $ & $            $ & $    1202 $ & $            $ \\
\hline
\end{tabular}
\label{tab:semi_pf_yields}
\end{table}

\begin{figure}[!hbtp]
\centering
\includegraphics[width=0.49\textwidth]{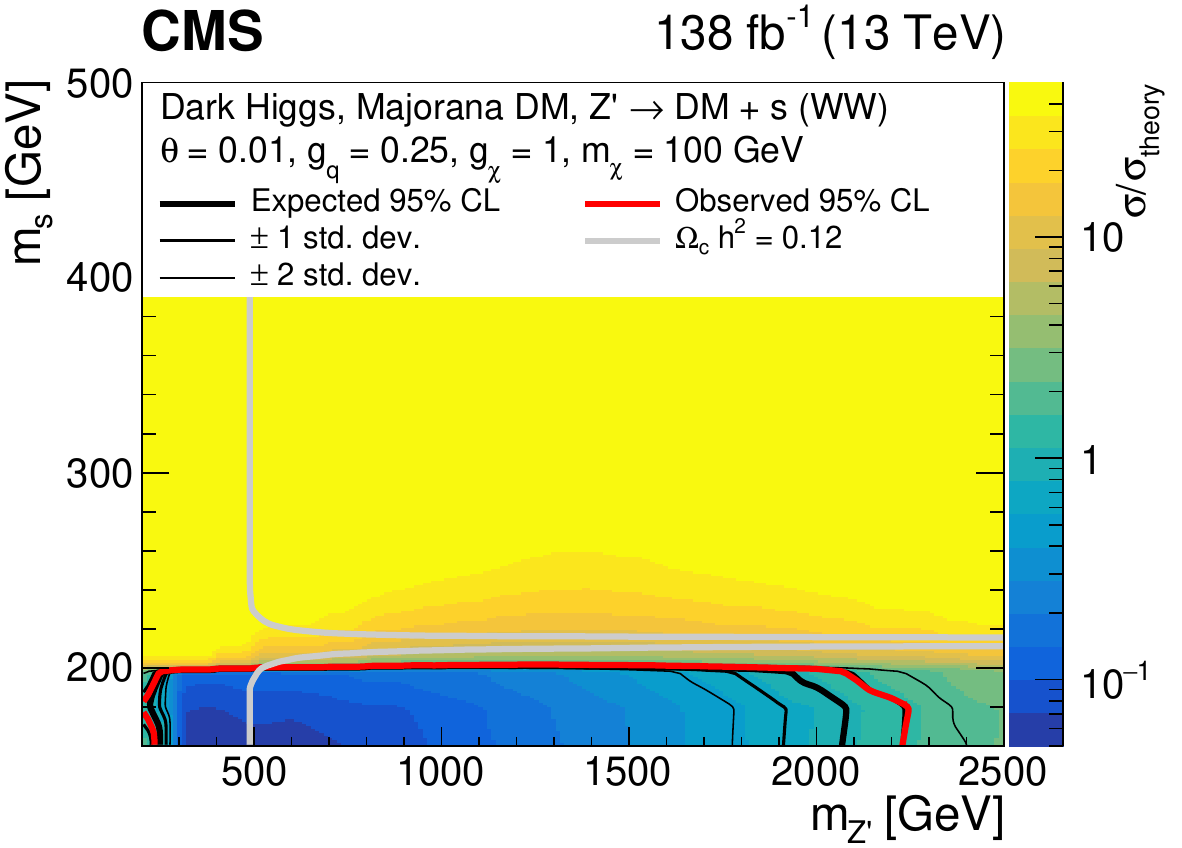}
\includegraphics[width=0.49\textwidth]{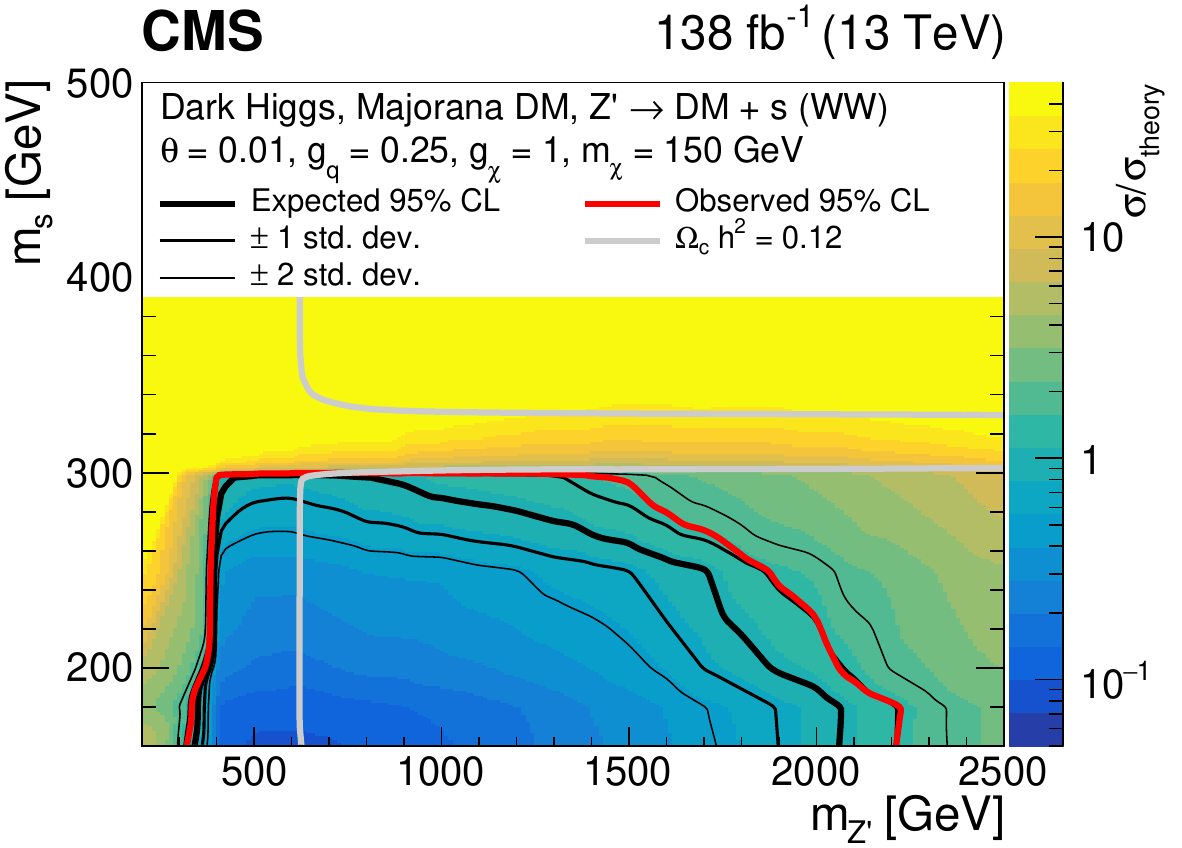}
\includegraphics[width=0.49\textwidth]{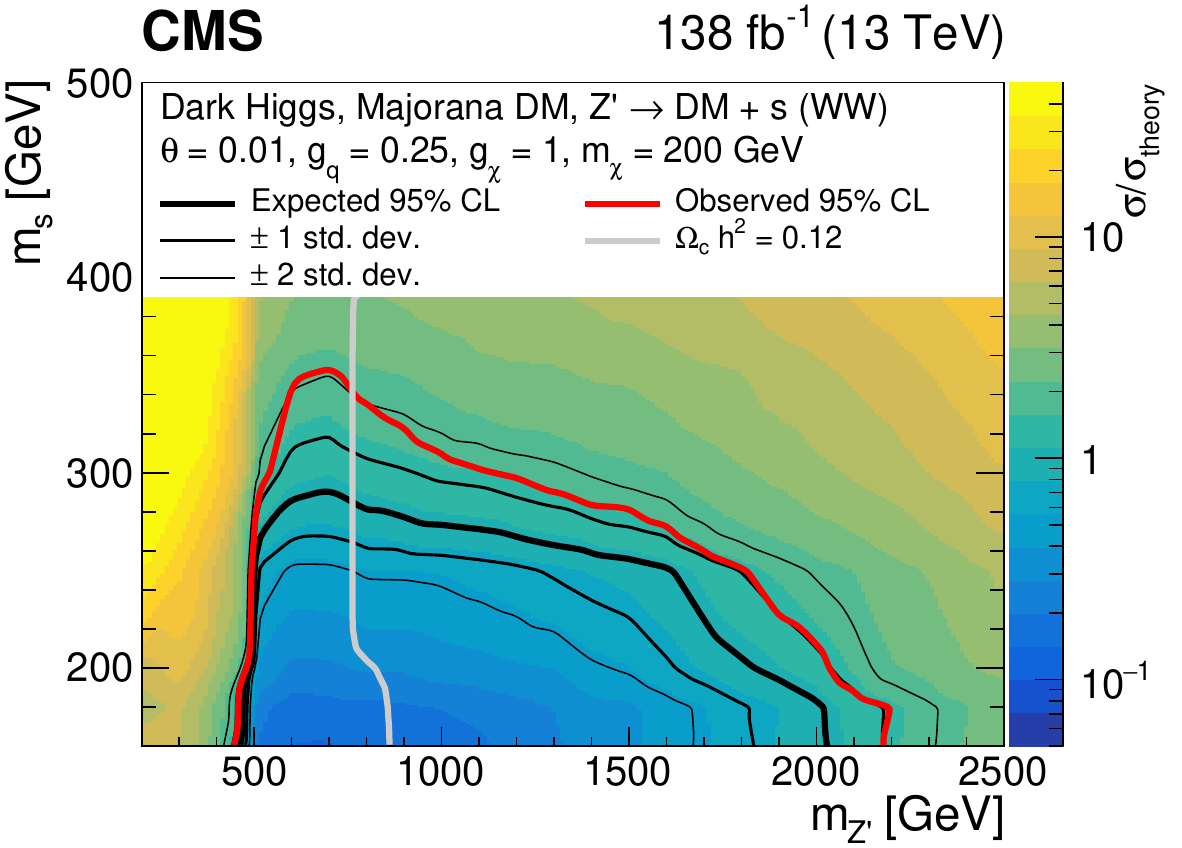}
\includegraphics[width=0.49\textwidth]{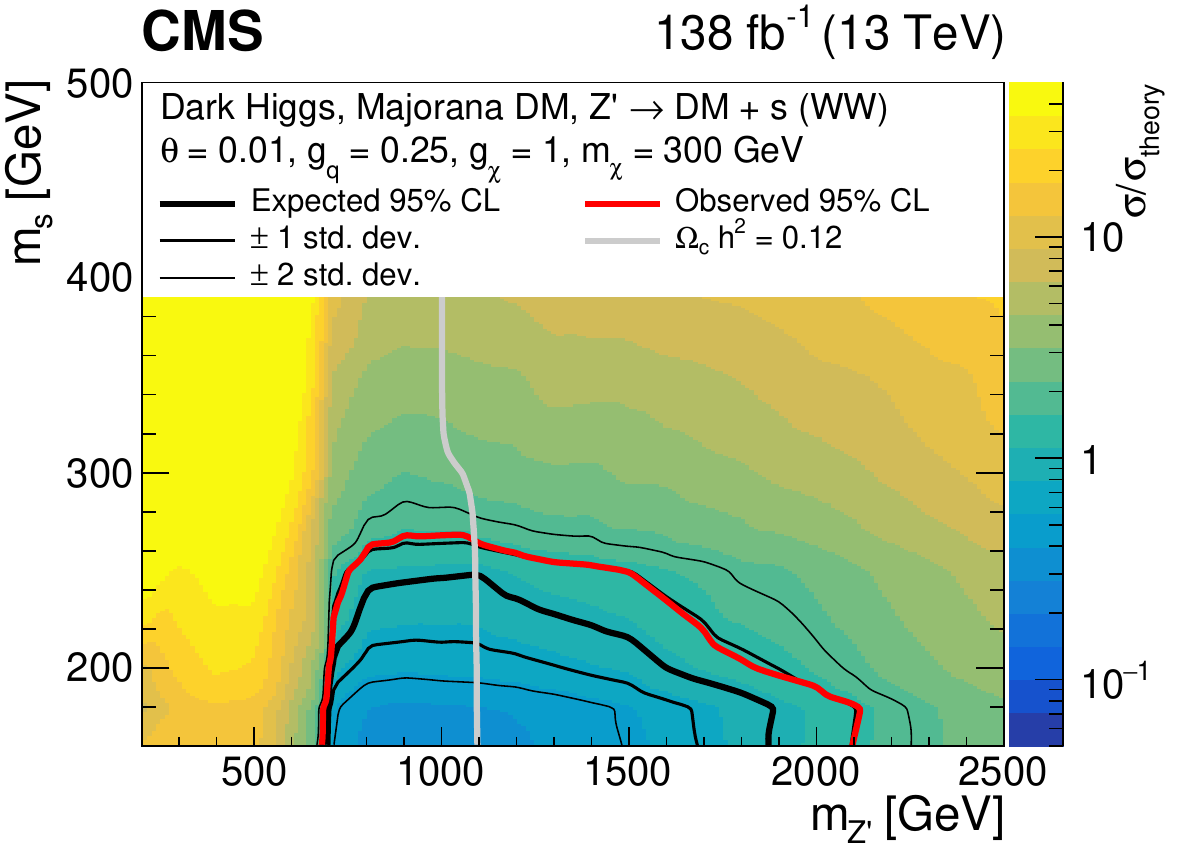}
\caption{\label{fig:darkHiggs_limit_comb}
  Observed (expected) exclusion regions at 95\% \CL for the dark Higgs model in the ($m_{\Hs}$, $m_{\PZpr}$) plane,
  marked by the solid red (black) line. The expected $\pm 1\sigma$ (68\% \CL) and $\pm 2\sigma$ (95\% \CL) bands 
  are shown as the thinner black lines.
  The bar on the righthand side of each figure maps the displayed colors to the corresponding limit values.
  Upper left: $m_{\chi} = 100\GeV$, upper right: $m_{\chi} = 150\GeV$,
  lower left: $m_{\chi} = 200\GeV$, lower right: $m_{\chi} = 300\GeV$.
  The gray line indicates were the model parameters produce exactly the observed relic density $\Omega_{\mathrm{c}} h^2 = 0.12$~\cite{ref:Planck}.}
\end{figure}

Neither the \Full nor the \Semi channel show a significant deviation from the SM predictions.
To extract upper limits, a modified frequentist approach was pursued using the \CLs criterion~\cite{Junk_1999,Read_2002}.
Upper limits at 95\% confidence level (\CL) on the model production cross section are obtained from the combination of
both final-states; these upper limits are displayed in the ($m_{\Hs}$, $m_{\PZpr}$) mass plane in Fig.~\ref{fig:darkHiggs_limit_comb}.
An interpolation between the signal samples is carried out by reweighting the events using the ratio of dark Higgs boson \pt; 
the reweighted samples are rescaled according to the respective theoretical cross sections to obtain the correct kinematic distributions.
A deficit of data events in the most sensitive bins, namely, the high-\mTlminmet bins in the \Full 
channel (with a local significance of 1.8 and 2.3  sigma in the highest \mll bin of SR1 and SR2 ) 
and the high-BDT bins in the \Semi channel (with a local significance of 1.8 sigma in the last SR bin of 2017--2018), 
leads to an observed limit that is more stringent than the expected one. 
The differences are not statistically significant, however, and the observed limit
falls within two standard deviations of the expected one for most of the scanned parameter space.
For $m_{\chi} = 200\GeV$, near $m_{\Hs} = 350\GeV$ and $m_{\PZpr} = 700\GeV$, 
the observed exclusion limit is just over two standard deviations above the expected one. 

Comparison with the observed DM relic density can indicate the preferred model parameters.
Therefore, relic density calculations are performed with the current dark Higgs model assumptions using \MADDM~\cite{ref:maddm}.
When the dark Higgs boson mass is lower than the DM mass, DM annihilation to two \Hs bosons is on-shell, 
thereby reducing the relic density compared to models with only the \PZpr boson as the mediator.
In the case where $m_{\Hs} \approx 2 m_{\chi}$, the WIMPs can be converted to SM particles through an on-shell dark Higgs resonance, 
strongly reducing the relic density.
Gray lines in Fig.~\ref{fig:darkHiggs_limit_comb} indicate where the model parameters produce exactly 
the current measurement of the observed relic density~\cite{ref:Planck}.

In this analysis, only the decay of the dark Higgs boson to a pair of visible \PW bosons is considered; this decay mode 
is dominant in the phase space analyzed. In the case where $m_{\Hs} > 2 m_{\chi}$, however, the dark Higgs boson decays
predominantly to a pair of DM particles. The consequence of this change of decay mode 
can be seen in Fig.~\ref{fig:darkHiggs_limit_comb}:
there is a boundary reflecting a sharp drop of sensitivity in the upper left (upper right) plot corresponding to
$m_{\Hs}$ equal to twice the DM particle mass of 100 (150)\GeV.

This search covers a wider range of DM mass (100–-300\GeV) compared to previous analyses~\cite{ref:ATLAS-DarkHiggs_1,ref:ATLAS-DarkHiggs_2}. 
In the ($m_{\Hs}$, $m_{\PZpr}$) plane with $m_{\chi} = 200\GeV$,
$m_{\PZpr}$ values up to ${\approx}2200\GeV$ are excluded for $m_{\Hs}$ close to $160\GeV$,
extending the lower limit on $m_{\PZpr}$ presented in Ref.~\cite{ref:ATLAS-DarkHiggs_2}.
In the same plane, the lower limit on $m_{\Hs}$ is ${\approx}350\GeV$ for $m_{\PZpr}$ around $700\GeV$,
which is slightly weaker than that of Ref.~\cite{ref:ATLAS-DarkHiggs_2}.

\section{Summary}\label{sec:summary}

A search for dark matter particles $\chi$ produced in association with a dark Higgs boson (\Hs) has been presented.
Proton-proton collision data at a center-of-mass energy of 13\TeV are used, corresponding to
an integrated luminosity of 138\fbinv. The decay mode of the dark Higgs boson to a $\PWp\PWm$ pair is explored.
Results are presented from a combination of the \Full and \Semi decay channels of the $\PWp\PWm$ pair (where $\Pell =$ electrons or muons).
No significant deviation from the standard model prediction is observed. Upper limits at 95\% confidence level
on the production cross section for dark matter particles are set and translated into bounds on dark Higgs model parameters.
This analysis investigates a dark matter mass range 100--300\GeV, which is wider than in previous searches
and extends the limit on the \PZpr boson mass $m_{\PZpr}$ in the region of the \Hs mass 
$160 < m_{\Hs} \lesssim 250\GeV$ for $m_{\chi} = 200\GeV$.
The most stringent limit is set for $m_{\chi} = 200\GeV$, excluding $m_{\Hs}$ masses up to ${\approx}350\GeV$
at $m_{\PZpr} = 700\GeV$, and up to $m_{\PZpr} \approx 2200\GeV$ for $m_{\Hs} = 160\GeV$.

\begin{acknowledgments}
We congratulate our colleagues in the CERN accelerator departments for the excellent performance of the LHC and thank the technical and administrative staffs at CERN and at other CMS institutes for their contributions to the success of the CMS effort. In addition, we gratefully acknowledge the computing centers and personnel of the Worldwide LHC Computing Grid and other centers for delivering so effectively the computing infrastructure essential to our analyses. Finally, we acknowledge the enduring support for the construction and operation of the LHC, the CMS detector, and the supporting computing infrastructure provided by the following funding agencies: SC (Armenia), BMBWF and FWF (Austria); FNRS and FWO (Belgium); CNPq, CAPES, FAPERJ, FAPERGS, and FAPESP (Brazil); MES and BNSF (Bulgaria); CERN; CAS, MoST, and NSFC (China); MINCIENCIAS (Colombia); MSES and CSF (Croatia); RIF (Cyprus); SENESCYT (Ecuador); MoER, ERC PUT and ERDF (Estonia); Academy of Finland, MEC, and HIP (Finland); CEA and CNRS/IN2P3 (France); SRNSF (Georgia); BMBF, DFG, and HGF (Germany); GSRI (Greece); NKFIH (Hungary); DAE and DST (India); IPM (Iran); SFI (Ireland); INFN (Italy); MSIP and NRF (Republic of Korea); MES (Latvia); LAS (Lithuania); MOE and UM (Malaysia); BUAP, CINVESTAV, CONACYT, LNS, SEP, and UASLP-FAI (Mexico); MOS (Montenegro); MBIE (New Zealand); PAEC (Pakistan); MES and NSC (Poland); FCT (Portugal); MESTD (Serbia); MCIN/AEI and PCTI (Spain); MOSTR (Sri Lanka); Swiss Funding Agencies (Switzerland); MST (Taipei); MHESI and NSTDA (Thailand); TUBITAK and TENMAK (Turkey); NASU (Ukraine); STFC (United Kingdom); DOE and NSF (USA).
  
\hyphenation{Rachada-pisek} Individuals have received support from the Marie-Curie program and the European Research Council and Horizon 2020 Grant, contract Nos.\ 675440, 724704, 752730, 758316, 765710, 824093, and COST Action CA16108 (European Union); the Leventis Foundation; the Alfred P.\ Sloan Foundation; the Alexander von Humboldt Foundation; the Science Committee, project no. 22rl-037 (Armenia); the Belgian Federal Science Policy Office; the Fonds pour la Formation \`a la Recherche dans l'Industrie et dans l'Agriculture (FRIA-Belgium); the Agentschap voor Innovatie door Wetenschap en Technologie (IWT-Belgium); the F.R.S.-FNRS and FWO (Belgium) under the ``Excellence of Science -- EOS" -- be.h project n.\ 30820817; the Beijing Municipal Science \& Technology Commission, No. Z191100007219010 and Fundamental Research Funds for the Central Universities (China); the Ministry of Education, Youth and Sports (MEYS) of the Czech Republic; the Shota Rustaveli National Science Foundation, grant FR-22-985 (Georgia); the Deutsche Forschungsgemeinschaft (DFG), under Germany's Excellence Strategy -- EXC 2121 ``Quantum Universe" -- 390833306, and under project number 400140256 - GRK2497; the Hellenic Foundation for Research and Innovation (HFRI), Project Number 2288 (Greece); the Hungarian Academy of Sciences, the New National Excellence Program - \'UNKP, the NKFIH research grants K 124845, K 124850, K 128713, K 128786, K 129058, K 131991, K 133046, K 138136, K 143460, K 143477, 2020-2.2.1-ED-2021-00181, and TKP2021-NKTA-64 (Hungary); the Council of Science and Industrial Research, India; ICSC -- National Research Center for High Performance Computing, Big Data and Quantum Computing, funded by the EU NexGeneration program (Italy); the Latvian Council of Science; the Ministry of Education and Science, project no. 2022/WK/14, and the National Science Center, contracts Opus 2021/41/B/ST2/01369 and 2021/43/B/ST2/01552 (Poland); the Funda\c{c}\~ao para a Ci\^encia e a Tecnologia, grant CEECIND/01334/2018 (Portugal); the National Priorities Research Program by Qatar National Research Fund; MCIN/AEI/10.13039/501100011033, ERDF ``a way of making Europe", and the Programa Estatal de Fomento de la Investigaci{\'o}n Cient{\'i}fica y T{\'e}cnica de Excelencia Mar\'{\i}a de Maeztu, grant MDM-2017-0765 and Programa Severo Ochoa del Principado de Asturias (Spain); the Chulalongkorn Academic into Its 2nd Century Project Advancement Project, and the National Science, Research and Innovation Fund via the Program Management Unit for Human Resources \& Institutional Development, Research and Innovation, grant B05F650021 (Thailand); the Kavli Foundation; the Nvidia Corporation; the SuperMicro Corporation; the Welch Foundation, contract C-1845; and the Weston Havens Foundation (USA).
\end{acknowledgments}

\bibliography{auto_generated}

\providecommand{\href}[2]{#2}\begingroup\raggedright\begin{thebibliography}{10}%
\makeatletter
\providecommand{\hrefCMSnoop }[0]{\@secondoftwo}%
\makeatother
\providecommand{\doi}{\texttt{doi:}\begingroup \urlstyle{tt}\Url}

\bibitem{Gaitskell:2004gd}
\hrefCMSnoop {}{R.~J. Gaitskell, ``Direct detection of dark matter'',} \textit{
  Ann. Rev. Nucl. Part. Sci.} \textbf{ 54} (2004) 315,
\href{http://dx.doi.org/10.1146/annurev.nucl.54.070103.181244}{\doi{10.1146/annurev.nucl.54.070103.181244}}.

\bibitem{Trimble:1987ee}
\hrefCMSnoop {}{V.~Trimble, ``Existence and nature of dark matter in the
  universe'',} \textit{ Ann. Rev. Astron. Astrophys.} \textbf{ 25} (1987) 425,
\href{http://dx.doi.org/10.1146/annurev.aa.25.090187.002233}{\doi{10.1146/annurev.aa.25.090187.002233}}.

\bibitem{Porter:2011nv}
\hrefCMSnoop {}{T.~A. Porter, R.~P. Johnson, and P.~W. Graham, ``Dark matter
  searches with astroparticle data'',} \textit{ Ann. Rev. Astron. Astrophys.}
  \textbf{ 49} (2011) 155,
  \href{http://dx.doi.org/10.1146/annurev-astro-081710-102528}{\doi{10.1146/annurev-astro-081710-102528}},
\href{http://www.arXiv.org/abs/1104.2836}{\texttt{arXiv:1104.2836}}.

\bibitem{ref:DMevi}
\hrefCMSnoop {}{G.~Bertone, D.~Hooper, and J.~Silk, ``Particle dark matter:
  evidence,candidates and constraints'',} \textit{ Phys. Rept.} \textbf{ 405}
  (2005) 279,
  \href{http://dx.doi.org/10.1016/j.physrep.2004.08.031}{\doi{10.1016/j.physrep.2004.08.031}},
\href{http://www.arXiv.org/abs/hep-ph/0404175}{\texttt{arXiv:hep-ph/0404175}}.

\bibitem{ref:DMNPBSM}
\hrefCMSnoop {}{J.~L. Feng, ``Dark matter candidates from particle physics and
  methods of detection'',} \textit{ Ann. Rev. Astron. Astrophys.} \textbf{ 48}
  (2010) 495,
  \href{http://dx.doi.org/10.1146/annurev-astro-082708-101659}{\doi{10.1146/annurev-astro-082708-101659}},
  \href{http://www.arXiv.org/abs/1003.0904}{\texttt{arXiv:1003.0904}}.

\bibitem{ref:relic}
\hrefCMSnoop {}{R.~J. Scherrer and M.~S. Turner, ``On the relic,cosmic
  abundance of stable,weakly interacting massive particles'',} \textit{ Phys.
  Rev. D} \textbf{ 33} (1986) 1585,
  \href{http://dx.doi.org/10.1103/PhysRevD.33.1585}{\doi{10.1103/PhysRevD.33.1585}}.

\bibitem{ref:Planck}
\hrefCMSnoop {}{{Planck} Collaboration, ``Planck 2018 results. {VI.}
  cosmological parameters'',} \textit{ Astron. Astrophys.} \textbf{ 641} (2020)
  A6,
  \href{http://dx.doi.org/10.1051/0004-6361/201833910}{\doi{10.1051/0004-6361/201833910}},
  \href{http://www.arXiv.org/abs/1807.06209}{\texttt{arXiv:1807.06209}}.
  [Erratum: \DOI{10.1051/0004-6361/201833910e}].

\bibitem{ref:WIMPs}
\hrefCMSnoop {}{G.~Steigman and M.~S. Turner, ``Cosmological constraints on the
  properties of weakly interacting massive particles'',} \textit{ Nucl. Phys.
  B} \textbf{ 253} (1985) 375,
\href{http://dx.doi.org/10.1016/0550-3213(85)90537-1}{\doi{10.1016/0550-3213(85)90537-1}}.

\bibitem{ref:monojet}
\hrefCMSnoop {}{{ATLAS Collaboration}, ``Search for dark matter and other new
  phenomena in events with an energetic jet and large missing transverse
  momentum using the {ATLAS} detector'',} \textit{ JHEP} \textbf{ 01} (2018)
  126,
  \href{http://dx.doi.org/10.1007/JHEP01(2018)126}{\doi{10.1007/JHEP01(2018)126}},
\href{http://www.arXiv.org/abs/1711.03301}{\texttt{arXiv:1711.03301}}.

\bibitem{ref:monojet1}
\hrefCMSnoop {}{{CMS Collaboration}, ``Search for new physics in final states
  with an energetic jet or a hadronically decaying {\PW} or {\PZ} boson and
  transverse momentum imbalance at $\sqrt{s}=13$ {TeV}'',} \textit{ Phys. Rev.
  D} \textbf{ 97} (2018) 092005,
  \href{http://dx.doi.org/10.1103/PhysRevD.97.092005}{\doi{10.1103/PhysRevD.97.092005}},
\href{http://www.arXiv.org/abs/1712.02345}{\texttt{arXiv:1712.02345}}.

\bibitem{ref:monob2}
\hrefCMSnoop {}{{ATLAS Collaboration}, ``Search for dark matter produced in
  association with bottom or top quarks in $\sqrt{s} = 13$ {TeV} pp collisions
  with the {ATLAS} detector'',} \textit{ Eur. Phys. J. C} \textbf{ 78} (2018)
  18,
  \href{http://dx.doi.org/10.1140/epjc/s10052-017-5486-1}{\doi{10.1140/epjc/s10052-017-5486-1}},
\href{http://www.arXiv.org/abs/1710.11412}{\texttt{arXiv:1710.11412}}.

\bibitem{ref:monob}
\hrefCMSnoop {}{{CMS Collaboration}, ``Search for dark matter in events with
  energetic,hadronically decaying top quarks and missing transverse momentum at
  $\sqrt{s}=13$ {TeV}'',} \textit{ JHEP} \textbf{ 06} (2018) 027,
  \href{http://dx.doi.org/10.1007/JHEP06(2018)027}{\doi{10.1007/JHEP06(2018)027}},
\href{http://www.arXiv.org/abs/1801.08427}{\texttt{arXiv:1801.08427}}.

\bibitem{ref:monog}
\hrefCMSnoop {}{{ATLAS Collaboration}, ``Search for dark matter at
  $\sqrt{s}=13$ {TeV} in final states containing an energetic photon and large
  missing transverse momentum with the {ATLAS} detector'',} \textit{ Eur. Phys.
  J. C} \textbf{ 77} (2017) 393,
  \href{http://dx.doi.org/10.1140/epjc/s10052-017-4965-8}{\doi{10.1140/epjc/s10052-017-4965-8}},
\href{http://www.arXiv.org/abs/1704.03848}{\texttt{arXiv:1704.03848}}.

\bibitem{ref:monog1}
\hrefCMSnoop {}{{CMS Collaboration}, ``Search for new physics in the monophoton
  final state in proton-proton collisions at $ \sqrt{s}=13 $ {TeV}'',} \textit{
  JHEP} \textbf{ 10} (2017) 073,
  \href{http://dx.doi.org/10.1007/JHEP10(2017)073}{\doi{10.1007/JHEP10(2017)073}},
\href{http://www.arXiv.org/abs/1706.03794}{\texttt{arXiv:1706.03794}}.

\bibitem{ref:monoW}
\hrefCMSnoop {}{{ATLAS Collaboration}, ``Search for an invisibly decaying higgs
  boson or dark matter candidates produced in association with a $z$ boson in
  pp collisions at $\sqrt{s} = 13$ {TeV} with the {ATLAS} detector'',} \textit{
  Phys. Lett. B} \textbf{ 776} (2018) 318,
  \href{http://dx.doi.org/10.1016/j.physletb.2017.11.049}{\doi{10.1016/j.physletb.2017.11.049}},
\href{http://www.arXiv.org/abs/1708.09624}{\texttt{arXiv:1708.09624}}.

\bibitem{Sirunyan:2017qfc}
\hrefCMSnoop {}{{CMS Collaboration}, ``Search for new physics in events with a
  leptonically decaying {\PZ} boson and a large transverse momentum imbalance
  in proton-proton collisions at $\sqrt{s} $ = 13 $ {TeV}$'',} \textit{ Eur.
  Phys. J. C} \textbf{ 78} (2018) 291,
  \href{http://dx.doi.org/10.1140/epjc/s10052-018-5740-1}{\doi{10.1140/epjc/s10052-018-5740-1}},
\href{http://www.arXiv.org/abs/1711.00431}{\texttt{arXiv:1711.00431}}.

\bibitem{ref:monoW1}
\hrefCMSnoop {}{{ATLAS Collaboration}, ``Search for dark matter in events with
  a hadronically decaying vector boson and missing transverse momentum in pp
  collisions at $\sqrt{s} = 13$ {TeV} with the {ATLAS} detector'',} \textit{
  JHEP} \textbf{ 10} (2018) 180,
  \href{http://dx.doi.org/10.1007/JHEP10(2018)180}{\doi{10.1007/JHEP10(2018)180}},
\href{http://www.arXiv.org/abs/1807.11471}{\texttt{arXiv:1807.11471}}.

\bibitem{Sirunyan_2020}
\hrefCMSnoop {}{{CMS Collaboration}, ``Search for dark matter particles
  produced in association with a {Higgs} boson in proton-proton collisions at
  $\sqrt{s} = 13$ {TeV}'',} \textit{ JHEP} \textbf{ 03} (2020) 25,
  \href{http://dx.doi.org/10.1007/jhep03(2020)025}{\doi{10.1007/jhep03(2020)025}},
  \href{http://www.arXiv.org/abs/1908.01713}{\texttt{arXiv:1908.01713}}.

\bibitem{Duerr_2017}
M.~Duerr\hrefCMSnoop {}{ {et~al.}, ``Hunting the dark {Higgs}'',} \textit{
  JHEP} \textbf{ 04} (2017) 143,
  \href{http://dx.doi.org/10.1007/jhep04(2017)143}{\doi{10.1007/jhep04(2017)143}},
  \href{http://www.arXiv.org/abs/1701.08780}{\texttt{arXiv:1701.08780}}.

\bibitem{WIMPSsafer}
M.~Duerr\hrefCMSnoop {}{ {et~al.}, ``How to save the {WIMP}: global analysis of
  a dark matter model with two $s$-channel mediators'',} \textit{ JHEP}
  \textbf{ 09} (2016) 42,
  \href{http://dx.doi.org/10.1007/JHEP09(2016)042}{\doi{10.1007/JHEP09(2016)042}},
  \href{http://www.arXiv.org/abs/1606.07609}{\texttt{arXiv:1606.07609}}.

\bibitem{Bell_2016}
\hrefCMSnoop {}{N.~F. Bell, Y.~Cai, and R.~K. Leane, ``Dark forces in the sky:
  signals from {Z'} and the dark {Higgs}'',} \textit{ JCAP} \textbf{ 08} (2016)
  001,
  \href{http://dx.doi.org/10.1088/1475-7516/2016/08/001}{\doi{10.1088/1475-7516/2016/08/001}},
  \href{http://www.arXiv.org/abs/1605.09382}{\texttt{arXiv:1605.09382}}.

\bibitem{ATL-PHYS-PUB-2019-032}
\href {http://cds.cern.ch/record/2686290}{{ATLAS Collaboration}, ``{RECAST}
  framework reinterpretation of an {ATLAS} dark matter search constraining a
  model of a dark {Higgs} boson decaying to two $b$-quarks'',} technical
  report, CERN, ATLAS Notes,ATL-PHYS-PUB-2019-032,2019.

\bibitem{ref:ATLAS-DarkHiggs_1}
\hrefCMSnoop {}{{ATLAS Collaboration}, ``Search for dark matter produced in
  association with a dark {Higgs} boson decaying into {$\PWp\PWm$} or
  {$\PZ\PZ$} in fully hadronic final states from $\sqrt{s} = 13$ {TeV} $pp$
  collisions recorded with the {ATLAS} detector'',} \textit{ Phys. Rev. Lett.}
  \textbf{ 126} (2021) 121802,
  \href{http://dx.doi.org/10.1103/PhysRevLett.126.121802}{\doi{10.1103/PhysRevLett.126.121802}},
  \href{http://www.arXiv.org/abs/2010.06548}{\texttt{arXiv:2010.06548}}.

\bibitem{ref:ATLAS-DarkHiggs_2}
\hrefCMSnoop {}{{ATLAS Collaboration}, ``Search for dark matter produced in
  association with a dark higgs boson decaying into $w^{+}w^{-}$ in the
  one-lepton final state at $\sqrt{s}$=13 {TeV} using 139 fb$^{-1}$ of $pp$
  collisions recorded with the {ATLAS} detector'',} \textit{ JHEP} \textbf{ 07}
  (2023) 116,
  \href{http://dx.doi.org/10.1007/JHEP07(2023)116}{\doi{10.1007/JHEP07(2023)116}},
  \href{http://www.arXiv.org/abs/2211.07175}{\texttt{arXiv:2211.07175}}.

\bibitem{hepdata}
\hrefCMSnoop {}{}{HEPD}ata record for this analysis, 2023.
\newblock
  \href{http://dx.doi.org/10.17182/hepdata.139719}{\doi{10.17182/hepdata.139719}}.

\bibitem{CMSTrigger}
\hrefCMSnoop {}{{CMS Collaboration}, ``The {CMS} trigger system'',} \textit{
  JINST} \textbf{ 12} (2017) P01020,
  \href{http://dx.doi.org/10.1088/1748-0221/12/01/P01020}{\doi{10.1088/1748-0221/12/01/P01020}},
\href{http://www.arXiv.org/abs/1609.02366}{\texttt{arXiv:1609.02366}}.

\bibitem{Chatrchyan:2008zzk}
\hrefCMSnoop {}{{CMS Collaboration}, ``The {CMS} experiment at the {CERN}
  {LHC}'',} \textit{ JINST} \textbf{ 3} (2008) S08004,
\href{http://dx.doi.org/10.1088/1748-0221/3/08/S08004}{\doi{10.1088/1748-0221/3/08/S08004}}.

\bibitem{CMS-LUM-17-003}
\hrefCMSnoop {}{{CMS Collaboration}, ``Precision luminosity measurement in
  proton-proton collisions at $\sqrt{s} =$ 13 {TeV} in 2015 and 2016 at
  {CMS}'',} \textit{ Eur. Phys. J. C} \textbf{ 81} (2021) 800,
  \href{http://dx.doi.org/10.1140/epjc/s10052-021-09538-2}{\doi{10.1140/epjc/s10052-021-09538-2}},
  \href{http://www.arXiv.org/abs/2104.01927}{\texttt{arXiv:2104.01927}}.

\bibitem{CMS-PAS-LUM-17-004}
\href {https://cds.cern.ch/record/2621960}{{CMS Collaboration}, ``{CMS}
  luminosity measurement for the 2017 data-taking period at $\sqrt{s} = 13$
  {TeV}'',} Technical Report CMS-PAS-LUM-17-004, CERN, CMS Physics Analysis
  Summary,CMS-PAS-LUM-17-004,2018.

\bibitem{CMS-PAS-LUM-18-002}
\href {https://cds.cern.ch/record/2676164}{{CMS Collaboration}, ``{CMS}
  luminosity measurement for the 2018 data-taking period at $\sqrt{s} = 13$
  {TeV}'',} Technical Report CMS-PAS-LUM-18-002, CERN, CMS Physics Analysis
  Summary,CMS-PAS-LUM-18-002,2019.

\bibitem{Sjostrand:2014zea}
T.~Sj{\"o}strand\hrefCMSnoop {}{ {et~al.}, ``An introduction to {PYTHIA}
  8.2'',} \textit{ Comput. Phys. Commun.} \textbf{ 191} (2015) 159,
  \href{http://dx.doi.org/10.1016/j.cpc.2015.01.024}{\doi{10.1016/j.cpc.2015.01.024}},
\href{http://www.arXiv.org/abs/1410.3012}{\texttt{arXiv:1410.3012}}.

\bibitem{Khachatryan:2015pea}
\hrefCMSnoop {}{{CMS Collaboration}, ``Event generator tunes obtained from
  underlying event and multiparton scattering measurements'',} \textit{ Eur.
  Phys. J. C} \textbf{ 76} (2016) 155,
  \href{http://dx.doi.org/10.1140/epjc/s10052-016-3988-x}{\doi{10.1140/epjc/s10052-016-3988-x}},
\href{http://www.arXiv.org/abs/1512.00815}{\texttt{arXiv:1512.00815}}.

\bibitem{Sirunyan:2019dfx}
\hrefCMSnoop {}{{CMS Collaboration}, ``Extraction and validation of a new set
  of {CMS} \textsc{PYTHIA}8 tunes from underlying-event measurements'',}
  \textit{ Eur. Phys. J. C} \textbf{ 80} (2020) 4,
  \href{http://dx.doi.org/10.1140/epjc/s10052-019-7499-4}{\doi{10.1140/epjc/s10052-019-7499-4}},
\href{http://www.arXiv.org/abs/1903.12179}{\texttt{arXiv:1903.12179}}.

\bibitem{Ball:2013hta}
\hrefCMSnoop {}{{NNPDF} Collaboration, ``Parton distributions with {QED}
  corrections'',} \textit{ Nucl. Phys. B} \textbf{ 877} (2013) 290,
  \href{http://dx.doi.org/10.1016/j.nuclphysb.2013.10.010}{\doi{10.1016/j.nuclphysb.2013.10.010}},
\href{http://www.arXiv.org/abs/1308.0598}{\texttt{arXiv:1308.0598}}.

\bibitem{Ball:2011uy}
\hrefCMSnoop {}{{NNPDF} Collaboration, ``Unbiased global determination of
  parton distributions and their uncertainties at {NNLO} and at {LO}'',}
  \textit{ Nucl. Phys. B} \textbf{ 855} (2012) 153,
  \href{http://dx.doi.org/10.1016/j.nuclphysb.2011.09.024}{\doi{10.1016/j.nuclphysb.2011.09.024}},
\href{http://www.arXiv.org/abs/1107.2652}{\texttt{arXiv:1107.2652}}.

\bibitem{Ball:2017nwa}
\hrefCMSnoop {}{{NNPDF} Collaboration, ``Parton distributions from
  high-precision collider data'',} \textit{ Eur. Phys. J. C} \textbf{ 77}
  (2017) 663,
  \href{http://dx.doi.org/10.1140/epjc/s10052-017-5199-5}{\doi{10.1140/epjc/s10052-017-5199-5}},
\href{http://www.arXiv.org/abs/1706.00428}{\texttt{arXiv:1706.00428}}.

\bibitem{Alwall_2014}
J.~Alwall\hrefCMSnoop {}{ {et~al.}, ``The automated computation of tree-level
  and next-to-leading order differential cross sections,and their matching to
  parton shower simulations'',} \textit{ JHEP} \textbf{ 07} (2014) 079,
  \href{http://dx.doi.org/10.1007/JHEP07(2014)079}{\doi{10.1007/JHEP07(2014)079}},
  \href{http://www.arXiv.org/abs/1405.0301}{\texttt{arXiv:1405.0301}}.

\bibitem{Abercrombie_2020}
D.~Abercrombie\hrefCMSnoop {}{ {et~al.}, ``Dark matter benchmark models for
  early {LHC Run-2} searches: {Report} of the {ATLAS/CMS} dark matter forum'',}
  \textit{ Phys. Dark Univ.} \textbf{ 27} (2020) 100371,
  \href{http://dx.doi.org/10.1016/j.dark.2019.100371}{\doi{10.1016/j.dark.2019.100371}},
  \href{http://www.arXiv.org/abs/1507.00966v1}{\texttt{arXiv:1507.00966v1}}.

\bibitem{Melia:2011tj}
\hrefCMSnoop {}{T.~Melia, P.~Nason, R.~R{\"o}ntsch, and G.~Zanderighi,
  ``{$\PW^{+}\PW^{-}$},{$\PW\PZ$} and {$\PZ\PZ$} production in the {POWHEG}
  {BOX}'',} \textit{ JHEP} \textbf{ 11} (2011) 078,
  \href{http://dx.doi.org/10.1007/JHEP11(2011)078}{\doi{10.1007/JHEP11(2011)078}},
  \href{http://www.arXiv.org/abs/1107.5051}{\texttt{arXiv:1107.5051}}.

\bibitem{Campbell:2011bn}
\hrefCMSnoop {}{J.~M. Campbell, R.~K. Ellis, and C.~Williams, ``Vector boson
  pair production at the {LHC}'',} \textit{ JHEP} \textbf{ 07} (2011) 018,
  \href{http://dx.doi.org/10.1007/JHEP07(2011)018}{\doi{10.1007/JHEP07(2011)018}},
\href{http://www.arXiv.org/abs/1105.0020}{\texttt{arXiv:1105.0020}}.

\bibitem{Campbell:2015qma}
\hrefCMSnoop {}{J.~M. Campbell, R.~K. Ellis, and W.~T. Giele, ``A
  multi-threaded version of {MCFM}'',} \textit{ Eur. Phys. J. C} \textbf{ 75}
  (2015) 246,
  \href{http://dx.doi.org/10.1140/epjc/s10052-015-3461-2}{\doi{10.1140/epjc/s10052-015-3461-2}},
\href{http://www.arXiv.org/abs/1503.06182}{\texttt{arXiv:1503.06182}}.

\bibitem{Meade:2014fc}
\hrefCMSnoop {}{P.~Meade, H.~Ramani, and M.~Zeng, ``Transverse momentum
  resummation effects in $\mathrm{W}^+\mathrm{W}^-$ measurements'',} \textit{
  Phys. Rev. D} \textbf{ 90} (2014) 114006,
  \href{http://dx.doi.org/10.1103/PhysRevD.90.114006}{\doi{10.1103/PhysRevD.90.114006}},
\href{http://www.arXiv.org/abs/1407.4481}{\texttt{arXiv:1407.4481}}.

\bibitem{Jaiswal:2014yba}
\hrefCMSnoop {}{P.~Jaiswal and T.~Okui, ``Explanation of the ww excess at the
  {LHC} by jet-veto resummation'',} \textit{ Phys. Rev. D} \textbf{ 90} (2014)
  073009,
  \href{http://dx.doi.org/10.1103/PhysRevD.90.073009}{\doi{10.1103/PhysRevD.90.073009}},
\href{http://www.arXiv.org/abs/1407.4537}{\texttt{arXiv:1407.4537}}.

\bibitem{gieseke2014vectorboson}
\hrefCMSnoop {}{S.~Gieseke, T.~Kasprzik, and J.~H. K{\"u}hn, ``Vector-boson
  pair production and electroweak corrections in {HERWIG++}'',} \textit{ Eur.
  Phys. J. C} \textbf{ 74} (2014) 2988,
  \href{http://dx.doi.org/10.1140/epjc/s10052-014-2988-y}{\doi{10.1140/epjc/s10052-014-2988-y}},
  \href{http://www.arXiv.org/abs/1401.3964}{\texttt{arXiv:1401.3964}}.

\bibitem{Caola:2015rqy}
\hrefCMSnoop {}{F.~Caola, K.~Melnikov, R.~R{\"o}ntsch, and L.~Tancredi, ``{QCD}
  corrections to $\mathrm{W}^+\mathrm{W}^-$ production through gluon fusion'',}
  \textit{ Phys. Lett. B} \textbf{ 754} (2016) 275,
  \href{http://dx.doi.org/10.1016/j.physletb.2016.01.046}{\doi{10.1016/j.physletb.2016.01.046}},
\href{http://www.arXiv.org/abs/1511.08617}{\texttt{arXiv:1511.08617}}.

\bibitem{Nason:2004rx}
\hrefCMSnoop {}{P.~Nason, ``A new method for combining {NLO} {QCD} with shower
  {Monte} {Carlo} algorithms'',} \textit{ JHEP} \textbf{ 11} (2004) 040,
  \href{http://dx.doi.org/10.1088/1126-6708/2004/11/040}{\doi{10.1088/1126-6708/2004/11/040}},
\href{http://www.arXiv.org/abs/hep-ph/0409146}{\texttt{arXiv:hep-ph/0409146}}.

\bibitem{Frixione:2007vw}
\hrefCMSnoop {}{S.~Frixione, P.~Nason, and C.~Oleari, ``Matching {NLO} {QCD}
  computations with parton shower simulations: the {POWHEG} method'',} \textit{
  JHEP} \textbf{ 11} (2007) 070,
  \href{http://dx.doi.org/10.1088/1126-6708/2007/11/070}{\doi{10.1088/1126-6708/2007/11/070}},
\href{http://www.arXiv.org/abs/0709.2092}{\texttt{arXiv:0709.2092}}.

\bibitem{Alioli:2010xd}
\hrefCMSnoop {}{S.~Alioli, P.~Nason, C.~Oleari, and E.~Re, ``A general
  framework for implementing {NLO} calculations in shower {Monte Carlo}
  programs: the {POWHEG BOX}'',} \textit{ JHEP} \textbf{ 06} (2010) 043,
  \href{http://dx.doi.org/10.1007/JHEP06(2010)043}{\doi{10.1007/JHEP06(2010)043}},
\href{http://www.arXiv.org/abs/1002.2581}{\texttt{arXiv:1002.2581}}.

\bibitem{Bagnaschi:2011tu}
\hrefCMSnoop {}{E.~Bagnaschi, G.~Degrassi, P.~Slavich, and A.~Vicini, ``{Higgs}
  production via gluon fusion in the {POWHEG} approach in the {SM} and in the
  {MSSM}'',} \textit{ JHEP} \textbf{ 02} (2012) 088,
  \href{http://dx.doi.org/10.1007/JHEP02(2012)088}{\doi{10.1007/JHEP02(2012)088}},
  \href{http://www.arXiv.org/abs/1111.2854}{\texttt{arXiv:1111.2854}}.

\bibitem{Nason:2009ai}
\hrefCMSnoop {}{P.~Nason and C.~Oleari, ``{NLO} {Higgs} boson production via
  vector-boson fusion matched with shower in {POWHEG}'',} \textit{ JHEP}
  \textbf{ 02} (2010) 037,
  \href{http://dx.doi.org/10.1007/JHEP02(2010)037}{\doi{10.1007/JHEP02(2010)037}},
\href{http://www.arXiv.org/abs/0911.5299}{\texttt{arXiv:0911.5299}}.

\bibitem{Luisoni:2013kna}
\hrefCMSnoop {}{G.~Luisoni, P.~Nason, C.~Oleari, and F.~Tramontano,
  ``{$\PH{}\PW^{\pm}{}/{}\PH{}\PZ$} + 0 and 1 jet at {NLO} with the {POWHEG
  BOX} interfaced to {GoSam} and their merging within {MiNLO}'',} \textit{
  JHEP} \textbf{ 10} (2013) 083,
  \href{http://dx.doi.org/10.1007/JHEP10(2013)083}{\doi{10.1007/JHEP10(2013)083}},
\href{http://www.arXiv.org/abs/1306.2542}{\texttt{arXiv:1306.2542}}.

\bibitem{Hartanto:2015uka}
\hrefCMSnoop {}{H.~B. Hartanto, B.~Jager, L.~Reina, and D.~Wackeroth, ``{Higgs}
  boson production in association with top quarks in the {POWHEG} {BOX}'',}
  \textit{ Phys. Rev. D} \textbf{ 91} (2015) 094003,
  \href{http://dx.doi.org/10.1103/PhysRevD.91.094003}{\doi{10.1103/PhysRevD.91.094003}},
\href{http://www.arXiv.org/abs/1501.04498}{\texttt{arXiv:1501.04498}}.

\bibitem{Bolognesi:2012mm}
S.~Bolognesi\hrefCMSnoop {}{ {et~al.}, ``On the spin and parity of a
  single-produced resonance at the {LHC}'',} \textit{ Phys. Rev. D} \textbf{
  86} (2012) 095031,
  \href{http://dx.doi.org/10.1103/PhysRevD.86.095031}{\doi{10.1103/PhysRevD.86.095031}},
\href{http://www.arXiv.org/abs/1208.4018}{\texttt{arXiv:1208.4018}}.

\bibitem{Czakon_2017}
M.~Czakon\hrefCMSnoop {}{ {et~al.}, ``Top-pair production at the {LHC} through
  {NNLO} {QCD} and {NLO} {EW}'',} \textit{ JHEP} \textbf{ 10} (2017) 186,
  \href{http://dx.doi.org/10.1007/jhep10(2017)186}{\doi{10.1007/jhep10(2017)186}},
  \href{http://www.arXiv.org/abs/1705.04105}{\texttt{arXiv:1705.04105}}.

\bibitem{Lindert:2017}
\hrefCMSnoop {}{J.~M. Lindert {et~al.}, ``Precise predictions for {\PV}+jets
  dark matter backgrounds'',} \textit{ Eur. Phys. J. C} \textbf{ 77} (2017)
  829,
  \href{http://dx.doi.org/10.1140/epjc/s10052-017-5389-1}{\doi{10.1140/epjc/s10052-017-5389-1}},
  \href{http://www.arXiv.org/abs/1705.04664}{\texttt{arXiv:1705.04664}}.

\bibitem{Melnikov:2006kv}
\hrefCMSnoop {}{K.~Melnikov and F.~Petriello, ``Electroweak gauge boson
  production at hadron colliders through $o(\alpha_s^2)$'',} \textit{ Phys.
  Rev. D} \textbf{ 74} (2006) 114017,
  \href{http://dx.doi.org/10.1103/PhysRevD.74.114017}{\doi{10.1103/PhysRevD.74.114017}},
\href{http://www.arXiv.org/abs/hep-ph/0609070}{\texttt{arXiv:hep-ph/0609070}}.

\bibitem{Agostinelli:2002hh}
\hrefCMSnoop {}{{GEANT4} Collaboration, ``{\GEANTfour}---a simulation
  toolkit'',} \textit{ Nucl. Instrum. Meth. A} \textbf{ 506} (2003) 250,
\href{http://dx.doi.org/10.1016/S0168-9002(03)01368-8}{\doi{10.1016/S0168-9002(03)01368-8}}.

\bibitem{CMS-TDR-15-02}
\href {http://cds.cern.ch/record/2020886}{{CMS Collaboration}, ``Technical
  proposal for the {Phase-II} upgrade of the {Compact Muon Solenoid}'',} CMS
  Technical Proposal CERN-LHCC-2015-010,CMS-TDR-15-02, 2015.

\bibitem{CMS-PRF-14-001}
\hrefCMSnoop {}{{CMS Collaboration}, ``Particle-flow reconstruction and global
  event description with the {CMS} detector'',} \textit{ JINST} \textbf{ 12}
  (2017) P10003,
  \href{http://dx.doi.org/10.1088/1748-0221/12/10/P10003}{\doi{10.1088/1748-0221/12/10/P10003}},
\href{http://www.arXiv.org/abs/1706.04965}{\texttt{arXiv:1706.04965}}.

\bibitem{Khachatryan:2015hwa}
\hrefCMSnoop {}{{CMS Collaboration}, ``Performance of electron reconstruction
  and selection with the {CMS} detector in proton-proton collisions at
  $\sqrt{s} = 8$ {TeV}'',} \textit{ JINST} \textbf{ 10} (2015) P06005,
  \href{http://dx.doi.org/10.1088/1748-0221/10/06/P06005}{\doi{10.1088/1748-0221/10/06/P06005}},
\href{http://www.arXiv.org/abs/1502.02701}{\texttt{arXiv:1502.02701}}.

\bibitem{Sirunyan:2018fpa}
\hrefCMSnoop {}{{CMS Collaboration}, ``Performance of the {CMS} muon detector
  and muon reconstruction with proton-proton collisions at
  {$\sqrt{s}=13\TeV$}'',} \textit{ JINST} \textbf{ 13} (2018) P06015,
  \href{http://dx.doi.org/10.1088/1748-0221/13/06/P06015}{\doi{10.1088/1748-0221/13/06/P06015}},
\href{http://www.arXiv.org/abs/1804.04528}{\texttt{arXiv:1804.04528}}.

\bibitem{Sirunyan:2019yvv}
\hrefCMSnoop {}{{CMS Collaboration}, ``Performance of the reconstruction and
  identification of high-momentum muons in proton-proton collisions at
  {$\sqrt{s}=13\TeV$}'',} \textit{ JINST} \textbf{ 15} (2020) P02027,
  \href{http://dx.doi.org/10.1088/1748-0221/15/02/P02027}{\doi{10.1088/1748-0221/15/02/P02027}},
\href{http://www.arXiv.org/abs/1912.03516}{\texttt{arXiv:1912.03516}}.

\bibitem{Cacciari:2008gp}
\hrefCMSnoop {}{M.~Cacciari, G.~P. Salam, and G.~Soyez, ``The anti-\kt jet
  clustering algorithm'',} \textit{ JHEP} \textbf{ 04} (2008) 063,
  \href{http://dx.doi.org/10.1088/1126-6708/2008/04/063}{\doi{10.1088/1126-6708/2008/04/063}},
  \href{http://www.arXiv.org/abs/0802.1189}{\texttt{arXiv:0802.1189}}.

\bibitem{Cacciari:2011ma}
\hrefCMSnoop {}{M.~Cacciari, G.~P. Salam, and G.~Soyez, ``{FastJet} user
  manual'',} \textit{ Eur. Phys. J. C} \textbf{ 72} (2012) 1896,
  \href{http://dx.doi.org/10.1140/epjc/s10052-012-1896-2}{\doi{10.1140/epjc/s10052-012-1896-2}},
\href{http://www.arXiv.org/abs/1111.6097}{\texttt{arXiv:1111.6097}}.

\bibitem{CMS-PAS-JME-16-003}
\href {https://cds.cern.ch/record/2256875}{{CMS Collaboration}, ``Jet
  algorithms performance in 13 {TeV} data'',} technical report, CERN, CMS
  Physics Analysis Summary,CMS-PAS-JME-16-003,2017.

\bibitem{Khachatryan_2017}
\hrefCMSnoop {}{{CMS Collaboration}, ``Jet energy scale and resolution in the
  {CMS} experiment in {$\Pp\Pp$} collisions at {8\TeV}'',} \textit{ JINST}
  \textbf{ 12} (2017) P02014,
  \href{http://dx.doi.org/10.1088/1748-0221/12/02/P02014}{\doi{10.1088/1748-0221/12/02/P02014}},
\href{http://www.arXiv.org/abs/1607.03663}{\texttt{arXiv:1607.03663}}.

\bibitem{Sirunyan:2017ezt}
\hrefCMSnoop {}{{CMS Collaboration}, ``Identification of heavy-flavour jets
  with the {CMS} detector in {\Pp{}\Pp} collisions at {13\TeV}'',} \textit{
  JINST} \textbf{ 13} (2018) P05011,
  \href{http://dx.doi.org/10.1088/1748-0221/13/05/P05011}{\doi{10.1088/1748-0221/13/05/P05011}},
\href{http://www.arXiv.org/abs/1712.07158}{\texttt{arXiv:1712.07158}}.

\bibitem{Sirunyan:2019kia}
\hrefCMSnoop {}{{CMS Collaboration}, ``Performance of missing transverse
  momentum reconstruction in proton-proton collisions at {$\sqrt{s} = 13\TeV$}
  using the {CMS} detector'',} \textit{ JINST} \textbf{ 14} (2019) P07004,
  \href{http://dx.doi.org/10.1088/1748-0221/14/07/P07004}{\doi{10.1088/1748-0221/14/07/P07004}},
\href{http://www.arXiv.org/abs/1903.06078}{\texttt{arXiv:1903.06078}}.

\bibitem{Bertolini:2014bba}
\hrefCMSnoop {}{D.~Bertolini, P.~Harris, M.~Low, and N.~Tran, ``Pileup per
  particle identification'',} \textit{ JHEP} \textbf{ 10} (2014) 059,
  \href{http://dx.doi.org/10.1007/JHEP10(2014)059}{\doi{10.1007/JHEP10(2014)059}},
\href{http://www.arXiv.org/abs/1407.6013}{\texttt{arXiv:1407.6013}}.

\bibitem{Moneta:2010pm}
L.~Moneta\hrefCMSnoop {}{ {et~al.}, ``The {RooStats} project'',} \textit{ PoS}
  \textbf{ ACAT2010} (2010) 057,
  \href{http://dx.doi.org/10.22323/1.093.0057}{\doi{10.22323/1.093.0057}},
  \href{http://www.arXiv.org/abs/1009.1003}{\texttt{arXiv:1009.1003}}.

\bibitem{Cowan2010AsymptoticFF}
\hrefCMSnoop {}{G.~Cowan, K.~Cranmer, E.~Gross, and O.~Vitells, ``Asymptotic
  formulae for likelihood-based tests of new physics'',} \textit{ Eur. Phys. J.
  C} \textbf{ 71} (2011) 1554,
  \href{http://dx.doi.org/10.1140/epjc/s10052-011-1554-0}{\doi{10.1140/epjc/s10052-011-1554-0}},
  \href{http://www.arXiv.org/abs/1007.1727}{\texttt{arXiv:1007.1727}}.
  [Erratum: \DOI{10.1140/epjc/s10052-013-2501-z}].

\bibitem{Sirunyan:2018egh}
\hrefCMSnoop {}{{CMS Collaboration}, ``Measurements of properties of the
  {Higgs} boson decaying to a {\PW} boson pair in {\Pp{}\Pp} collisions at
  {$\sqrt{s}=13\TeV$}'',} \textit{ Phys. Lett. B} \textbf{ 791} (2019) 96,
  \href{http://dx.doi.org/10.1016/j.physletb.2018.12.073}{\doi{10.1016/j.physletb.2018.12.073}},
\href{http://www.arXiv.org/abs/1806.05246}{\texttt{arXiv:1806.05246}}.

\bibitem{Sirunyan:2018nqx}
\hrefCMSnoop {}{{CMS Collaboration}, ``Measurement of the inelastic
  proton-proton cross section at {$\sqrt{s}=13\TeV$}'',} \textit{ JHEP}
  \textbf{ 07} (2018) 161,
  \href{http://dx.doi.org/10.1007/JHEP07(2018)161}{\doi{10.1007/JHEP07(2018)161}},
\href{http://www.arXiv.org/abs/1802.02613}{\texttt{arXiv:1802.02613}}.

\bibitem{Butterworth:2015oua}
\hrefCMSnoop {}{J.~Butterworth {et~al.}, ``{PDF4LHC} recommendations for {LHC
  Run II}'',} \textit{ J. Phys. G} \textbf{ 43} (2016) 023001,
  \href{http://dx.doi.org/10.1088/0954-3899/43/2/023001}{\doi{10.1088/0954-3899/43/2/023001}},
  \href{http://www.arXiv.org/abs/1510.03865}{\texttt{arXiv:1510.03865}}.

\bibitem{CMS:2022uhn}
\hrefCMSnoop {}{{CMS Collaboration}, ``Measurements of the higgs boson
  production cross section and couplings in the w boson pair decay channel in
  proton-proton collisions at $\sqrt{s}=13$ {TeV}'',} \textit{ Eur. Phys. J. C}
  \textbf{ 83} (2023) 667,
  \href{http://dx.doi.org/10.1140/epjc/s10052-023-11632-6}{\doi{10.1140/epjc/s10052-023-11632-6}},
  \href{http://www.arXiv.org/abs/2206.09466}{\texttt{arXiv:2206.09466}}.

\bibitem{Caola:2016trd}
F.~Caola\hrefCMSnoop {}{ {et~al.}, ``{QCD} corrections to vector boson pair
  production in gluon fusion including interference effects with off-shell
  {Higgs} at the {LHC}'',} \textit{ JHEP} \textbf{ 07} (2016) 087,
  \href{http://dx.doi.org/10.1007/JHEP07(2016)087}{\doi{10.1007/JHEP07(2016)087}},
  \href{http://www.arXiv.org/abs/1605.04610}{\texttt{arXiv:1605.04610}}.

\bibitem{BARLOW1993219}
\hrefCMSnoop {}{R.~Barlow and C.~Beeston, ``Fitting using finite {Monte}
  {Carlo} samples'',} \textit{ Computer Physics Communications} \textbf{ 77}
  (1993) 219,
  \href{http://dx.doi.org/10.1016/0010-4655(93)90005-W}{\doi{10.1016/0010-4655(93)90005-W}}.

\bibitem{Junk_1999}
\hrefCMSnoop {}{T.~Junk, ``Confidence level computation for combining searches
  with small statistics'',} \textit{ Nucl. Instrum. Meth. A} \textbf{ 434}
  (1999) 435,
  \href{http://dx.doi.org/10.1016/s0168-9002(99)00498-2}{\doi{10.1016/s0168-9002(99)00498-2}}.

\bibitem{Read_2002}
\hrefCMSnoop {}{A.~L. Read, ``Presentation of search results: the {\CLs}
  technique'',} \textit{ J. Phys. G} \textbf{ 28} (2002) 2693,
  \href{http://dx.doi.org/10.1088/0954-3899/28/10/313}{\doi{10.1088/0954-3899/28/10/313}}.

\bibitem{ref:maddm}
C.~Arina\hrefCMSnoop {}{ {et~al.}, ``Indirect dark-matter detection with {MadDM
  v3.2 \textendash{}} lines and loops'',} \textit{ Eur. Phys. J. C} \textbf{
  83} (2023) 241,
  \href{http://dx.doi.org/10.1140/epjc/s10052-023-11377-2}{\doi{10.1140/epjc/s10052-023-11377-2}},
  \href{http://www.arXiv.org/abs/2107.04598}{\texttt{arXiv:2107.04598}}.

\end{thebibliography}\endgroup
\cleardoublepage \appendix\section{The CMS Collaboration \label{app:collab}}\begin{sloppypar}\hyphenpenalty=5000\widowpenalty=500\clubpenalty=5000
\cmsinstitute{Yerevan Physics Institute, Yerevan, Armenia}
{\tolerance=6000
A.~Hayrapetyan, A.~Tumasyan\cmsAuthorMark{1}\cmsorcid{0009-0000-0684-6742}
\par}
\cmsinstitute{Institut f\"{u}r Hochenergiephysik, Vienna, Austria}
{\tolerance=6000
W.~Adam\cmsorcid{0000-0001-9099-4341}, J.W.~Andrejkovic, T.~Bergauer\cmsorcid{0000-0002-5786-0293}, S.~Chatterjee\cmsorcid{0000-0003-2660-0349}, K.~Damanakis\cmsorcid{0000-0001-5389-2872}, M.~Dragicevic\cmsorcid{0000-0003-1967-6783}, A.~Escalante~Del~Valle\cmsorcid{0000-0002-9702-6359}, P.S.~Hussain\cmsorcid{0000-0002-4825-5278}, M.~Jeitler\cmsAuthorMark{2}\cmsorcid{0000-0002-5141-9560}, N.~Krammer\cmsorcid{0000-0002-0548-0985}, D.~Liko\cmsorcid{0000-0002-3380-473X}, I.~Mikulec\cmsorcid{0000-0003-0385-2746}, J.~Schieck\cmsAuthorMark{2}\cmsorcid{0000-0002-1058-8093}, R.~Sch\"{o}fbeck\cmsorcid{0000-0002-2332-8784}, D.~Schwarz\cmsorcid{0000-0002-3821-7331}, M.~Sonawane\cmsorcid{0000-0003-0510-7010}, S.~Templ\cmsorcid{0000-0003-3137-5692}, W.~Waltenberger\cmsorcid{0000-0002-6215-7228}, C.-E.~Wulz\cmsAuthorMark{2}\cmsorcid{0000-0001-9226-5812}
% ...
\par}
\vskip\cmsinstskip
\dag:~Deceased\\
$^{1}$Also at Yerevan State University, Yerevan, Armenia\\
$^{2}$Also at TU Wien, Vienna, Austria\\
% ...
\end{sloppypar}
\end{document}